\documentclass[twocolumn]{aastex631}

\newcommand{\pixedfit}{\texttt{piXedfit}}

\usepackage{soul}  

\begin{document}

\title{Stellar Mass Assembly History of Massive Quiescent Galaxies since $z\sim4$: Insights from Spatially Resolved SED Fitting with JWST Data}

\author[0009-0009-3404-5673]{Novan Saputra Haryana}
\affiliation{Astronomical Institute, Tohoku University, 6-3, Aramaki, Aoba-ku, Sendai, Miyagi, 980-8578, Japan}

\author[0000-0002-2651-1701]{Masayuki Akiyama}
\affiliation{Astronomical Institute, Tohoku University, 6-3, Aramaki, Aoba-ku, Sendai, Miyagi, 980-8578, Japan}

\author[0000-0002-5258-8761]{Abdurro'uf}
\affiliation{Department of Physics and Astronomy, The Johns Hopkins University, 3400 N Charles St., Baltimore, MD 21218, USA}
\affiliation{Space Telescope Science Institute (STScI), 3700 San Martin Drive, Baltimore, MD 21218, USA}

\author[0000-0003-1551-519X]{Hesti Retno Tri Wulandari}
\affiliation{Astronomy Research Division, Institut Teknologi Bandung, Jl. Ganesha No. 10 Bandung 40132, Indonesia}
\affiliation{Bosscha Observatory, Institut Teknologi Bandung, Jl. Peneropongan Bintang, Lembang, Bandung Barat 40391 , Indonesia}

\author[0009-0007-3423-1332]{Juan Pablo Alfonzo}
\affiliation{Astronomical Institute, Tohoku University, 6-3, Aramaki, Aoba-ku, Sendai, Miyagi, 980-8578, Japan}

\author[0000-0003-4814-0101]{Kianhong Lee}
\affiliation{Astronomical Institute, Tohoku University, 6-3, Aramaki, Aoba-ku, Sendai, Miyagi, 980-8578, Japan}

\author[0000-0002-8299-0006]{Naoki Matsumoto}
\affiliation{Astronomical Institute, Tohoku University, 6-3, Aramaki, Aoba-ku, Sendai, Miyagi, 980-8578, Japan}

\author[0009-0005-1487-7772]{Ryo Albert Sutanto}
\affiliation{Astronomical Institute, Tohoku University, 6-3, Aramaki, Aoba-ku, Sendai, Miyagi, 980-8578, Japan}

\author[0009-0006-8034-6061]{Muhammad Nur Ihsan Effendi}
\affiliation{, Jl. Ganesha No. 10 Bandung 40132, Indonesia}

\author[0000-0002-5956-8018]{Itsna Khoirul Fitriana}
\affiliation{National Astronomical Observatory of Japan, 2-21-1, Osawa, Mitaka, Tokyo 181-8588, Japan}
\affiliation{Astronomy Research Division, Institut Teknologi Bandung, Jl. Ganesha No. 10 Bandung 40132, Indonesia}
\affiliation{Bosscha Observatory, Institut Teknologi Bandung, Jl. Peneropongan Bintang, Lembang, Bandung Barat 40391 , Indonesia}

\author[0000-0002-5897-2158]{Ibnu Nurul Huda}
\affiliation{School of Astronomy and Space Science, Key Laboratory of Modern Astronomy and Astrophysics (Ministry of Education), Nanjing University, No. 163 Xianlin Avenue, 210023 Nanjing, People's Republic of China}
\affiliation{Research Center for Computation, National Research and Innovation Agency, Jl. Sangkuriang No. 10, 40135 Bandung, Indonesia}

\author[0000-0001-6282-5778]{Anton Timur Jaelani}
\affiliation{Astronomy Research Division, Institut Teknologi Bandung, Jl. Ganesha No. 10 Bandung 40132, Indonesia}
\affiliation{Bosscha Observatory, Institut Teknologi Bandung, Jl. Peneropongan Bintang, Lembang, Bandung Barat 40391 , Indonesia}

\author[0009-0008-0862-8966]{Sultan Hadi Kusuma}
\affiliation{Master’s Program in Astronomy, Institut Teknologi Bandung, Jl. Ganesha No. 10 Bandung 40132, Indonesia}

\author{Lucky Puspitarini}
\affiliation{Astronomy Research Division, Institut Teknologi Bandung, Jl. Ganesha No. 10 Bandung 40132, Indonesia}
\affiliation{Bosscha Observatory, Institut Teknologi Bandung, Jl. Peneropongan Bintang, Lembang, Bandung Barat 40391 , Indonesia}

\author[0000-0002-4752-128X]{Dian Puspita Triani}
\affiliation{Center for Astrophysics Harvard \& Smithshonian, 60 Garden Street, Cambridge, MA 02138, USA}



\begin{abstract}

Massive quiescent galaxies at high redshift show significantly more compact morphology than their local counterparts. To examine their internal structure across a wide redshift range and investigate potential redshift dependence, we performed spatially resolved SED fitting using \pixedfit\ software on massive $(\log(M_*/M_\odot)\sim11)$ quiescent galaxies at $0<z<4$ with public James Webb Space Telescope and Hubble Space Telescope imaging data from the Public Release Imaging for Extragalactic Research and the Cosmic Evolution Early Release Science Survey. We find that at $z \sim 3.5$, the half-mass radius is about 5.4 times smaller than at $z \sim 0.5$. This growth is driven by stellar mass buildup in the outskirts ($r > 4$ kpc), while the central regions ($r \sim 1$ kpc) remain largely unchanged, with stellar mass surface density similar to local quiescent galaxies. The estimated star formation rates are too low to explain the stellar mass growth, indicating an additional stellar mass accumulation process, such as mergers, is necessary. We parameterize the size–mass relation of the most massive galaxies in our sample as $\log(R_{e,mass}) \propto \alpha \log(M_*)$, and find $\alpha = 2.67^{+1.14}_{-1.17}$ for $z\lessapprox2$, consistent with growth dominated by minor mergers, and $\alpha = 0.91^{+0.20}_{-0.16}$ for $z\gtrapprox2$, consistent with growth dominated by major mergers. These results indicate that massive quiescent galaxies originate from compact quenched systems and grow through combinations of minor and major mergers.

\end{abstract}




\section{Introduction} \label{sec:intro}

Galaxies in the universe are generally divided into two broad categories based on their star formation activities and colors. Star-forming galaxies are typically blue and actively forming stars, while quiescent galaxies appear red and exhibit little to no ongoing star formation \citep[e.g.,][]{2004ApJ...608..752B, Baldry_2004, 2009ApJ...706L.173B, 2014MNRAS.440..889S}. Quiescent galaxies, in particular, are key to understand the processes that regulate galaxy evolution. These galaxies have already passed the peak of their star formation activity, reflecting an evolutionary phase where the bulk of their stellar mass was assembled earlier in cosmic history \citep[e.g.,][]{2013ApJ...770L..39W, 2016ApJ...832...79P, 2023A&A...677A.184W}. Even after the end of major star formation activity, these galaxies continue to evolve structurally, making them valuable tracers of the mechanisms governing galaxy assembly over time \citep[e.g.,][]{2010ApJ...709.1018V, vanderWel_2014}.

Observations have shown that massive quiescent galaxies (\(\log M_*/M_{\odot} > 11\)) at high redshift (\(z > 2\)) are significantly more compact than their lower-redshift counterparts. Their effective radius at \(z \sim 3\) is 3–4 times smaller than at $z\sim0$ \citep[e.g.,][]{Newman_2012, vanderWel_2014, 2021MNRAS.501.1028Y, 2024ApJ...972..134M, 2024ApJ...964..192I}. In contrast, star-forming galaxies at similar redshifts generally exhibit larger effective radii and show a more moderate evolution in size over the same period (e.g., $\sim4.3$ times larger for galaxies with $M_*\sim5\times10^{10}M_\odot$ at $z\sim3$), {most likely driven by} the continuous addition of stellar mass through ongoing star formation and disk growth \citep[e.g.,][]{vanderWel_2014, 2015ApJS..219...15S, 2018MNRAS.479.5083A}. This difference implies that once star formation ceases, quiescent galaxies may be more vulnerable to mechanisms such as mergers or expansion \citep[e.g.,][]{2009ApJ...697.1290B}, {that more substantially drive structural evolution}. Understanding the mechanisms behind this size evolution is crucial for reconstructing the full evolutionary pathways of massive galaxies.

Several scenarios have been proposed to account for this observed size evolution. One possibility is that the apparent size growth is due to a progenitor bias, where quiescent galaxies observed at later times are inherently different from those seen at higher redshifts. A recent study by \citet{2024ApJ...971...99C} found that young massive quiescent galaxies at \(z > 1\) are significantly more compact than older quiescent galaxies with similar mass and redshift. They argue that quenching at later cosmic times occurs through a more diverse set of pathways, while at higher redshifts, the quenching mechanisms more commonly produce compact remnants. However, multiple studies have shown that progenitor bias alone cannot fully explain the observed size evolution, suggesting that an actual physical growth mechanism must be responsible \citep[e.g.,][]{Carollo_2013, Fagioli_2016}.

A widely accepted hypothesis is that minor mergers drive the size growth of quiescent galaxies over time. Observational studies have found that the evolution of effective radius (\(R_e\)) and total stellar mass (\(M_*\)) follows a relation consistent with minor mergers, approximately \( R_e \sim M_*^2 \) \citep[e.g.,][]{2009ApJ...697.1290B, 2010ApJ...709.1018V, 2013MNRAS.429.2924H}. This scaling is expected if low-mass galaxies are accreted onto the outskirts of more massive systems, increasing their size without significantly affecting their core structure. In contrast, major mergers, which would lead to a milder increase in size (\( R_e \sim M_* \)), appear to be less frequent among quiescent galaxies at later cosmic times, making them an unlikely primary driver of growth \citep[e.g.,][]{Newman_2012, 2015ApJ...799..206B}.  A study by \citet{2023ApJ...956L..42S} found that approximately 30\% of the stellar mass in massive quiescent galaxies is contributed by minor mergers. They also reported that companion galaxies with mass ratios below 1:10 typically exhibit bluer colors. A more recent study by \citet{2024MNRAS.532.3604C} found that older quiescent galaxies typically exhibit negative metallicity gradients and flat age gradients, suggesting they commonly experience minor dry mergers with low-metallicity companions.

The minor merger-driven scenario is further supported by cosmological simulations \citep[e.g.,][]{2009ApJ...699L.178N, 2012ApJ...744...63O, 2015MNRAS.449..361W, 2016MNRAS.458.2371R}, which have provided theoretical backing for the observed size evolution. These simulations indicate that the primary mode of stellar mass growth evolves over time. At high redshift ($z > 2$), galaxies grow mainly through in-situ star formation, rapidly assembling mass via intense starburst episodes. After quenching, however, external accretion, particularly through mergers, becomes the dominant driver of mass growth, especially at lower redshifts ($z < 2$). This transition from internal to external growth is a key prediction of numerical simulations, which show that as star formation shuts down, galaxies continue to evolve structurally by accreting stars through minor mergers. These mergers preferentially deposit material in the outer regions, leading to a gradual flattening of stellar mass profiles and an increase in effective radius over time. Importantly, simulations also indicate that while the outskirts of these galaxies grow significantly, their core structures remain largely unchanged, implying that the core structure of massive quiescent galaxies is already in place at the time of quenching \citep[e.g.,][]{2008MNRAS.389..567C, 2009ApJ...699L.178N, 2010ApJ...725.2312O, 2013MNRAS.429.2924H}.  

Despite significant progress in understanding the size evolution of quiescent galaxies, the underlying physical mechanisms driving this growth remain debated. In particular, the spatially resolved evolution of stellar mass and star formation activity across different cosmic epochs has not been fully explored. With the advent of the James Webb Space Telescope 
\citep[\textit{JWST};][]{2023PASP..135f8001G, 2023PASP..135d8001R}, we now have the sensitivity and spatial resolution needed to probe the internal structures of high-redshift galaxies in unprecedented detail. In this study, we aim to examine the physical processes responsible for the size growth of massive quiescent galaxies (\(M_* > 10^{11} M_\odot\)) across cosmic time (\(0 < z < 4\)) by conducting spatially resolved spectral energy distribution (SED) fitting using \pixedfit\ \citep{2021ApJS..254...15A, 2022ascl.soft07033A} on deep JWST/NIRCam imaging. Through this analysis, we investigate how the internal distributions of stellar mass and star formation evolve with redshift and evaluate different growth scenarios such as minor mergers, adiabatic expansion, and inside-out quenching. We use the cosmological parameters $H_0=70$ km s$^{-1}$ Mpc$^{-1}$, $\Omega_M = 0.3$, and $\Omega_\Lambda= 0.7$ throughout this paper.
\begin{figure}
    \centering
    \includegraphics[width=\linewidth]{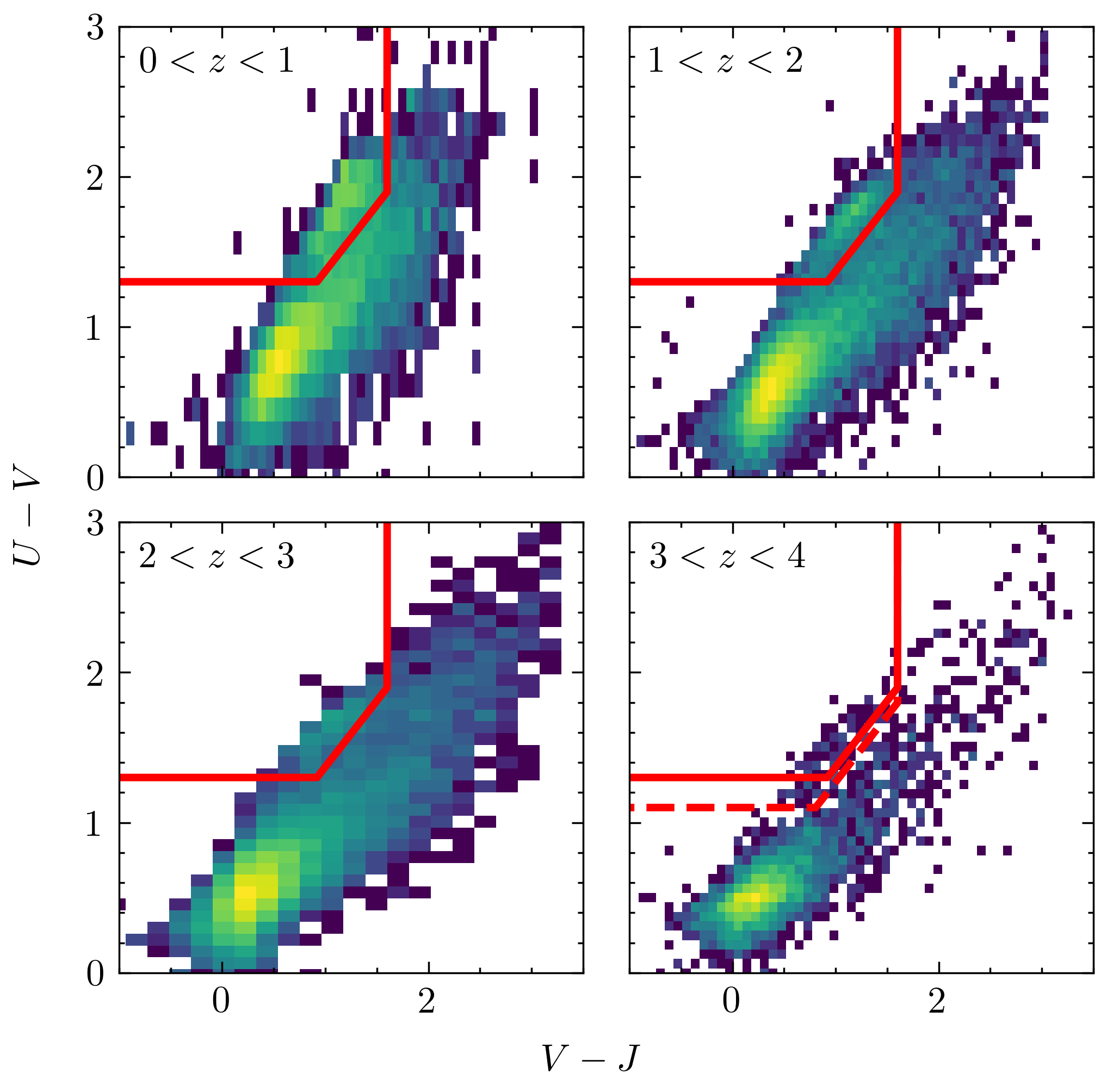}
    \caption{Rest-frame $U-V$ versus $V-J$ colors (i.e., \textit{UVJ} diagram) of galaxies in the DJA photometry catalog. Red solid lines are selection criteria for quiescent galaxies adopted from \cite{2009ApJ...691.1879W}. We extend our selection in the highest redshift bin (dashed red line) to include recently quenched galaxies, following \cite{2022A&A...666A.141M}. The sample selected with this extended quiescent box is used only in Sections \ref{sec:progen} and \ref{sec:discussion}.}
    \label{fig:eazyUVJ}
\end{figure}

\begin{figure}
    \centering
    \includegraphics[width=\linewidth]{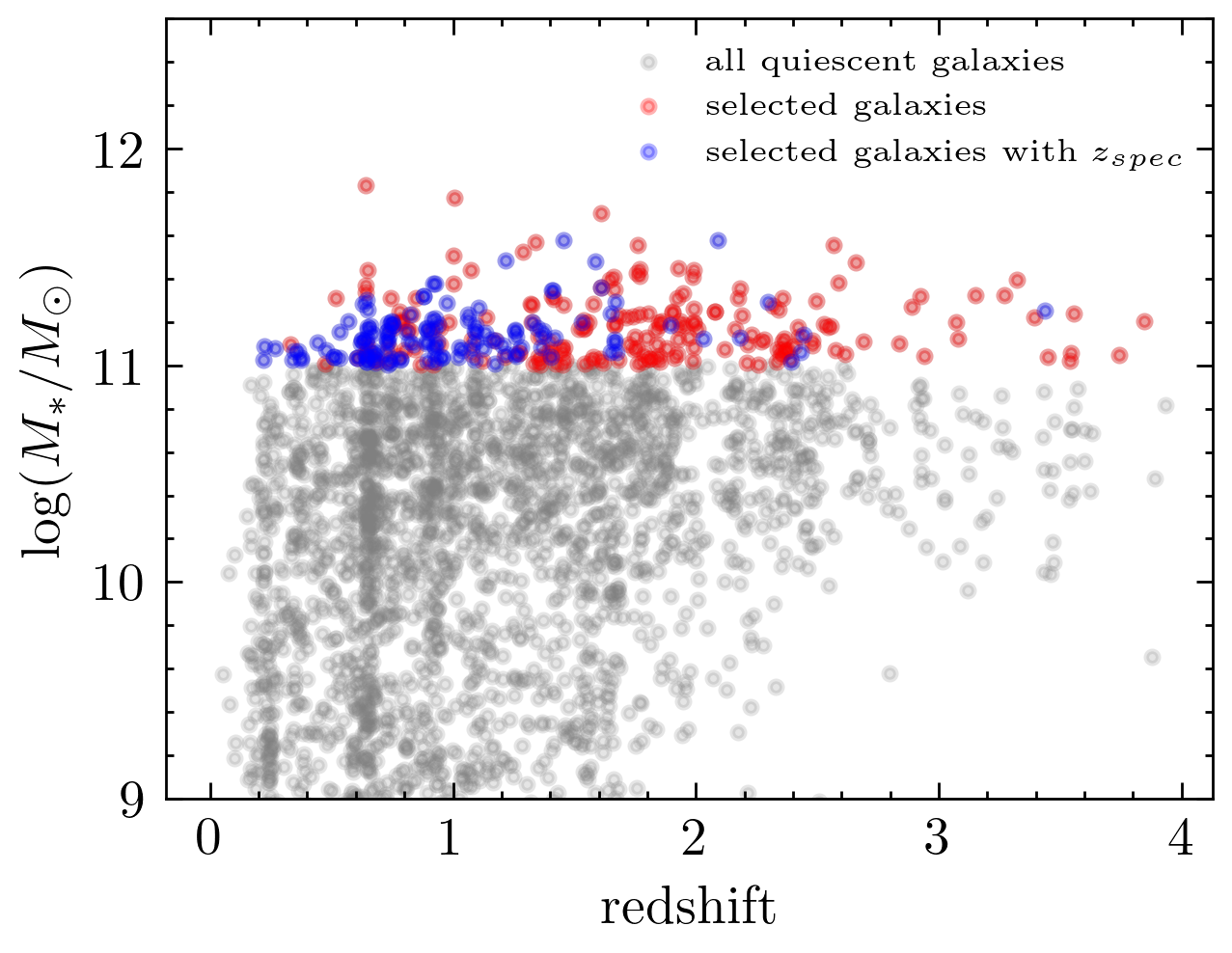}
    \caption{Stellar mass and redshift of quiescent galaxies with $z_{phot}<4$. Grey, red, and blue points show all quiescent galaxies, massive quiescent galaxies ($\log(M_*/M_\odot)>11$), and the massive quiescent galaxies with confirmed spectroscopic redshift, respectively.}
    \label{fig:mass_zphot}
\end{figure}
\section{Data and Samples} \label{sec:data_samples}

\subsection{Data}

\begin{table*}
    \centering
    \caption{Measured FWHM (milliarcseconds) of the empirical PSFs generated from the imaging data used in this study. For comparison, the two rightmost columns show estimated FWHM values for similar instruments from \cite{2024A&A...691A.240M} (ASTRODEEP) and theoretical PSFs computed with WebbPSF \citep{2015ascl.soft04007P}.}
    \begin{tabular}{l l c c c c c c c}
    \hline
        Telescope & Filter/Instrument & \multicolumn{7}{c}{PSF FWHM (milliarcsec)}  
        \\ 
        \hline
         &  & EGS & COSMOS & COSMOS & UDS & UDS & ASTRODEEP & WebbPSF\\ 
         & & & (west) & (east) & (south) & (north) & & \\
        \hline
        HST & ACS & & & \\
        \hline
        & F435W & 109 & 109 & 109 & 104 & 104 & 112 & -\\
        & F606W & 107 & 108 & 108 & 105 & 105 & 122 & -\\
        & F814W & 108 & 106 & 106 & 100 & 100 & 100 & -\\
        \hline
        JWST & NIRCam & & & \\
        \hline
        & F090W & - & 58 & 59 & 59 & 58 & 56 & 30\\
        & F115W & 62 & 58 & 58 & 58 & 60 & 59 & 37\\
        & F150W & 64 & 60 & 59 & 59 & 60 & 59 & 49\\
        & F200W & 79 & 70 & 71 & 72 & 70 & 73 & 64\\
        & F277W & 129 & 127 & 129 & 128 & 127 & 124 & 88\\
        & F356W & 148 & 146 & 147 & 148 & 149 & 146 & 114\\
        & F410M & 163 & 161 & 162 & 162 & 161 & 155 & 133\\
        & F444W & 172 & 168 & 170 & 169 & 168 & 166 & 140\\
        \hline
        \hline
    \end{tabular}
    \label{tab:psf}
\end{table*}

{We use publicly available JWST imaging data from three major extragalactic legacy fields: the Extended Groth Strip (EGS; ceers-full-grizli-v7.2), COSMOS (divided into primer-cosmos-west-grizli-v7.0 and primer-cosmos-east-grizli-v7.0), and the Ultra Deep Survey (UDS; divided into primer-uds-south-grizli-v7.2 and primer-uds-north-grizli-v7.2).} These fields have been extensively covered by the Hubble Space Telescope (\textit{\textit{HST}}) as part of the Cosmic Assembly Near-infrared Deep Extragalactic Legacy Survey (CANDELS; \citealt{2011ApJS..197...35G}, \citealt{2011ApJS..197...36K}), ensuring a robust set of complementary optical to near-infrared observations.

The JWST imaging data are taken from the DAWN JWST Archive (DJA)\footnote{\url{https://dawn-cph.github.io/dja/imaging/v7/}}, v7 version of the reductions. These data were processed using the \texttt{GRIZLI} package \citep{2023zndo...8370018B}, which drizzles them to a common pixel grid with a pixel scale of $0.^{\prime\prime}04$.  
For a detailed description of the data reduction procedures, please refer to \cite{2023ApJ...947...20V}.

We compile JWST data from several public programs. In the EGS field, we utilize data from the Cosmic Evolution Early Release Science Survey (CEERS; ERS 1345, PI Finkelstein; \citealt{2023ApJ...946L..13F}), combined with additional F444W imaging from GO 2279 (PI Naidu) and data from several other filters from GO 2750 (PI Arrabal-Haro). In total, we use JWST/NIRCam imaging data in seven filters, consisting of F115W, F150W, F200W, F277W, F356W, F410M, and F444W. For the UDS and COSMOS fields, the imaging data were taken from the Public Release Imaging for Extragalactic Research (PRIMER; GO 1837, PI Dunlop), which partially overlaps with the COSMOS-Web survey (GO 1727, PIs Kartaltepe \& Casey; \citealt{2023ApJ...954...31C}). In these fields, we make use of NIRCam data in eight filters, similar to those in EGS, but with the addition of the F090W band.

To supplement these JWST observations, we incorporate archival HST imaging data from the CANDELS survey. For all fields, we include HST/ACS images in the F435W, F606W, and F814W bands. These extensive multi-wavelength datasets are essential for placing strong constraints on the stellar population properties through SED fitting.

DJA also provides photometry catalogs along with basic global properties and redshift measurements derived from SED fitting. The photometry was performed using Source Extractor Python (\texttt{SEP}; \citealt{2016JOSS....1...58B}) and \texttt{GRIZLI}, while the SED fitting was performed using \texttt{eazy-py} \citep{2008ApJ...686.1503B, 2021zndo...7575984B} using the \texttt{agn\_blue\_sfhz\_13}\footnote{\url{https://github.com/gbrammer/eazy-photoz/blob/master/templates/sfhz/README.md}} template set. This template set is a set of star formation history templates that vary with redshift and are constrained to begin no earlier than the age of the universe at that epoch. The SED fitting incorporates additional MIRI JWST photometry. In addition to photometric redshift measurements, the catalogs also provide spectroscopic redshifts compiled from various sources in the literature. 

Throughout this work, we use spectroscopic redshifts when available, and photometric redshifts otherwise. {To ensure consistency, for galaxies with available spectroscopic redshifts, we re-fit their photometry using \texttt{eazy\-py} with the same \texttt{agn\_blue\_sfhz\_13} template set, fixing the redshift to the spectroscopic value.} To assess the reliability of the photometric redshifts, we compare predicted photometric redshifts with available spectroscopic measurements. We obtain that \( |z_{\text{spec}} - z_{\text{phot}}| / (1 + z_{\text{spec}}) \) has a median of \( 1.45 \times 10^{-2} \) with a normalized median absolute deviation (NMAD) of \( \sigma_{\text{NMAD}} \simeq 0.017 \). We use \( \sigma_{\text{NMAD}} \) as \citep[following][]{2008ApJ...686.1503B}  
\begin{equation}
    \sigma_{\text{NMAD}} = 1.48 \times \text{median} \left( \frac{|\Delta z - \text{median}(\Delta z)|}{1 + z_{\text{spec}}} \right),
\end{equation}  
where \( \Delta z = |z_{\text{spec}} - z_{\text{phot}}| \). Additionally, we find that only 7\% of galaxies meet the outlier criterion of \( |z_{\text{spec}} - z_{\text{phot}}| / (1 + z_{\text{spec}}) > 0.15 \). These indicate that the photometric redshift measurements are highly reliable.

\subsection{Initial Sample Selection}\label{sec:selections}

\begin{figure*}
    \centering
    \includegraphics[width=\linewidth]{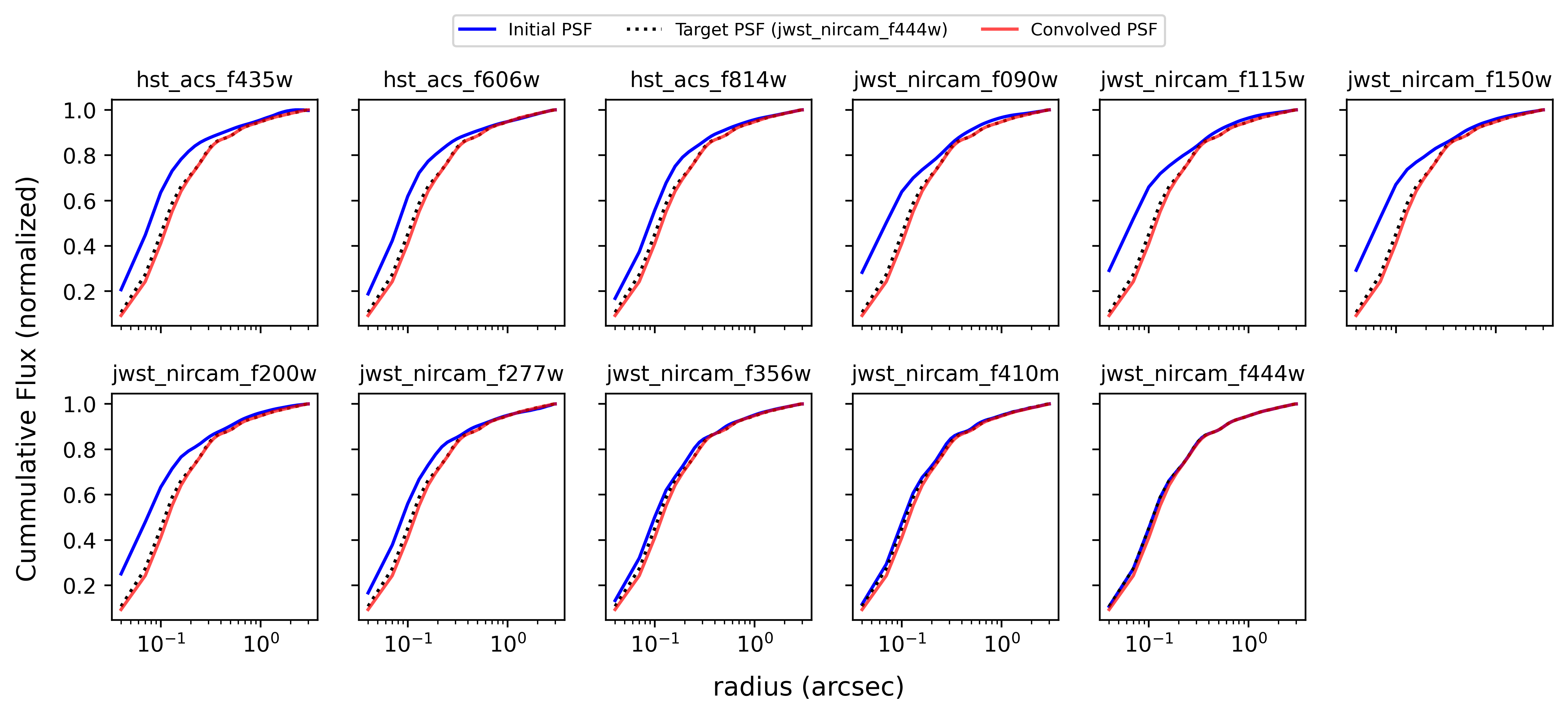}
    \caption{Curves of growth for the PSFs from primer-uds-south-grizli-v7.2 in each filter (blue), compared to the target PSF used for PSF matching (F444W; black dotted line). The red curves show the PSFs after convolution with the constructed kernels, demonstrating how closely they match the target PSF.}
    \label{fig:COG}
\end{figure*}

By using catalogs from DJA, we first select quiescent galaxies at $z<4$ based on the $UVJ$ diagram, following the criteria defined by \cite{2009ApJ...691.1879W}, as shown in Figure \ref{fig:eazyUVJ}. We then select massive galaxies based on stellar mass from the DJA catalogs, restricting the selection to $\log(M_*/M_\odot) > 11$. Our sample from these selections is illustrated in Figure \ref{fig:mass_zphot}. The selected galaxies span a redshift range between 0.23 and 3.85, within which the rest-frame $U$, $V$, and $J$ wavelengths are well covered by the photometric data. Based on these criteria, {we obtain a total of 330 galaxies, 140 of which have confirmed spectroscopic redshifts}.

\section{Methods} \label{sec:methods}

\subsection{Generating Empirical PSF and Convolution Kernels}

{To homogenize the spatial resolution across different filters in each field, we create convolution kernels for point spread function (PSF) matching. We construct empirical PSFs using the \texttt{PSFEx} software \citep{2011ASPC..442..435B} in conjunction with \texttt{SExtractor} \citep{1996A&AS..117..393B}. Point-like sources are first identified from the F200W band image using a customized selection method based on the \texttt{MU\_MAX} vs \texttt{MAG\_AUTO} diagnostic plane, as described in a DJA internship report by \cite{Genin2024} \citep[which was based on][]{2007ApJS..172..219L}. In this method, \texttt{MU\_MAX} (the peak surface brightness in mag/arcsec$^2$) and \texttt{MAG\_AUTO} (the total magnitude) are measured for each source using \texttt{SExtractor}. Stars form a tight linear sequence (“star-line”) in this plane, while extended sources (e.g., galaxies) form a more scattered cloud.}

{To isolate the star-line, we first remove the extended source cloud using DBSCAN \citep{1996kddm.conf..226E}, a density-based clustering algorithm implemented in \texttt{scikit-learn}. We then apply a robust linear regression using RANSAC \citep{10.1145/358669.358692} to fit the remaining star-line on the \texttt{MU\_MAX}-\texttt{MAG\_AUTO} vs \texttt{MAG\_AUTO} diagram. Point sources are then selected within a narrow box along this line, with width and magnitude range manually tuned to avoid saturation and contamination. Specifically, we adopt a \texttt{MAG\_AUTO} range of 19–28 mag and a vertical width of 0.8 mag in \texttt{MU\_MAX}–\texttt{MAG\_AUTO}. Using this method, the number of selected point sources in each field is as follows: primer-uds-south-grizli-v7.2 (490), primer-uds-north-grizli-v7.2 (438), primer-cosmos-west-grizli-v7.0 (326), primer-cosmos-east-grizli-v7.0 (368), and ceers-full-grizli-v7.2 (329).}

{After selecting these point sources, we extract their image vignettes and feed them to \texttt{PSFEx} to generate the empirical PSF. The FWHM of the resulting PSFs in each band is listed in Table \ref{tab:psf}. Although point-source selection is performed in the F200W band, the same stellar positions are used to derive PSFs in other bands, ensuring consistent sampling across filters. }

{The next step is to create convolution kernels for PSF matching. This process involves degrading the spatial resolution of all images to match that of the filter with the largest FWHM (F444W filter from NIRCam). The convolution kernel for matching a pair of PSFs is derived from the ratio of their Fourier transforms \citep[e.g.,][]{2008ApJ...682..336G, 2011PASP..123.1218A}. For details on the PSF matching procedure and its accuracy, see \cite{2021ApJS..254...15A} (Section 3.1.2.)\footnote{Details on the convolution kernels and demonstrations of their performance are available at \url{https://pixedfit.readthedocs.io/en/latest/list\_kernels\_psf.html}}. Figure \ref{fig:COG} shows examples of curves of growth for the PSFs before and after the matching procedure, for the field primer-uds-south-grizli-v7.2. After generating the matching kernels, we extract $150 \times 150$ pixel cutouts (corresponding to $6 \times 6$ arcsec) centered on each sample galaxy and convolve them with the appropriate kernels to achieve a uniform spatial resolution across all filters.}

\begin{figure}
    \centering
    \includegraphics[width=\linewidth]{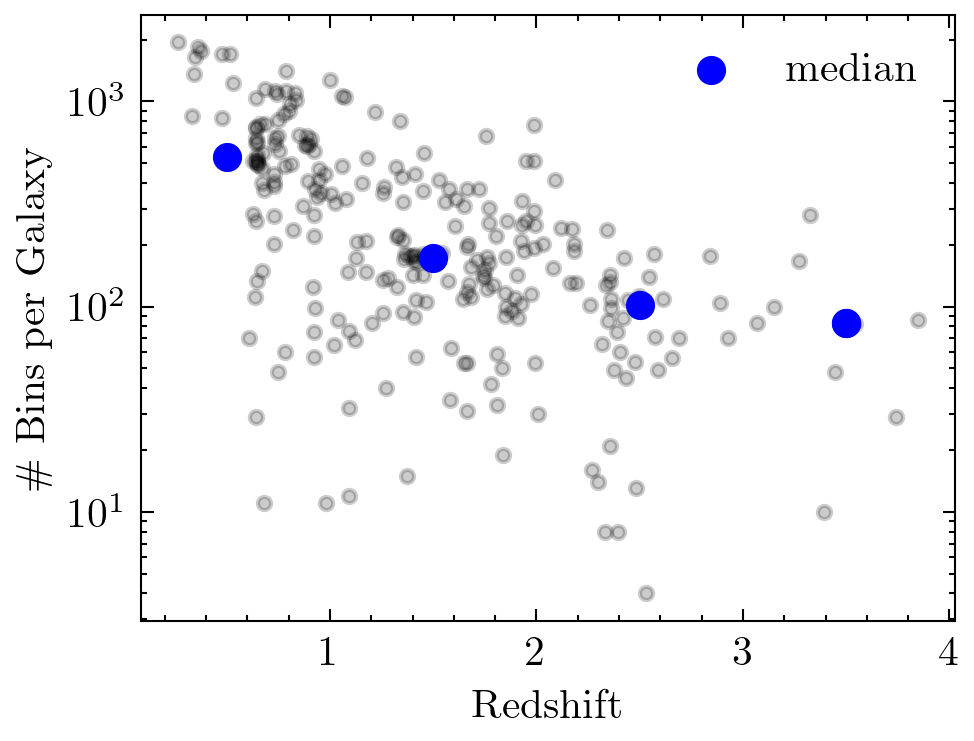}
    \caption{Number of bins per galaxy as a function of redshift. Blue points indicate the median values within each redshift bin.}
    \label{fig:bin_per_galaxy}
\end{figure}

\begin{figure*}
    \centering
    \includegraphics[width=\linewidth]{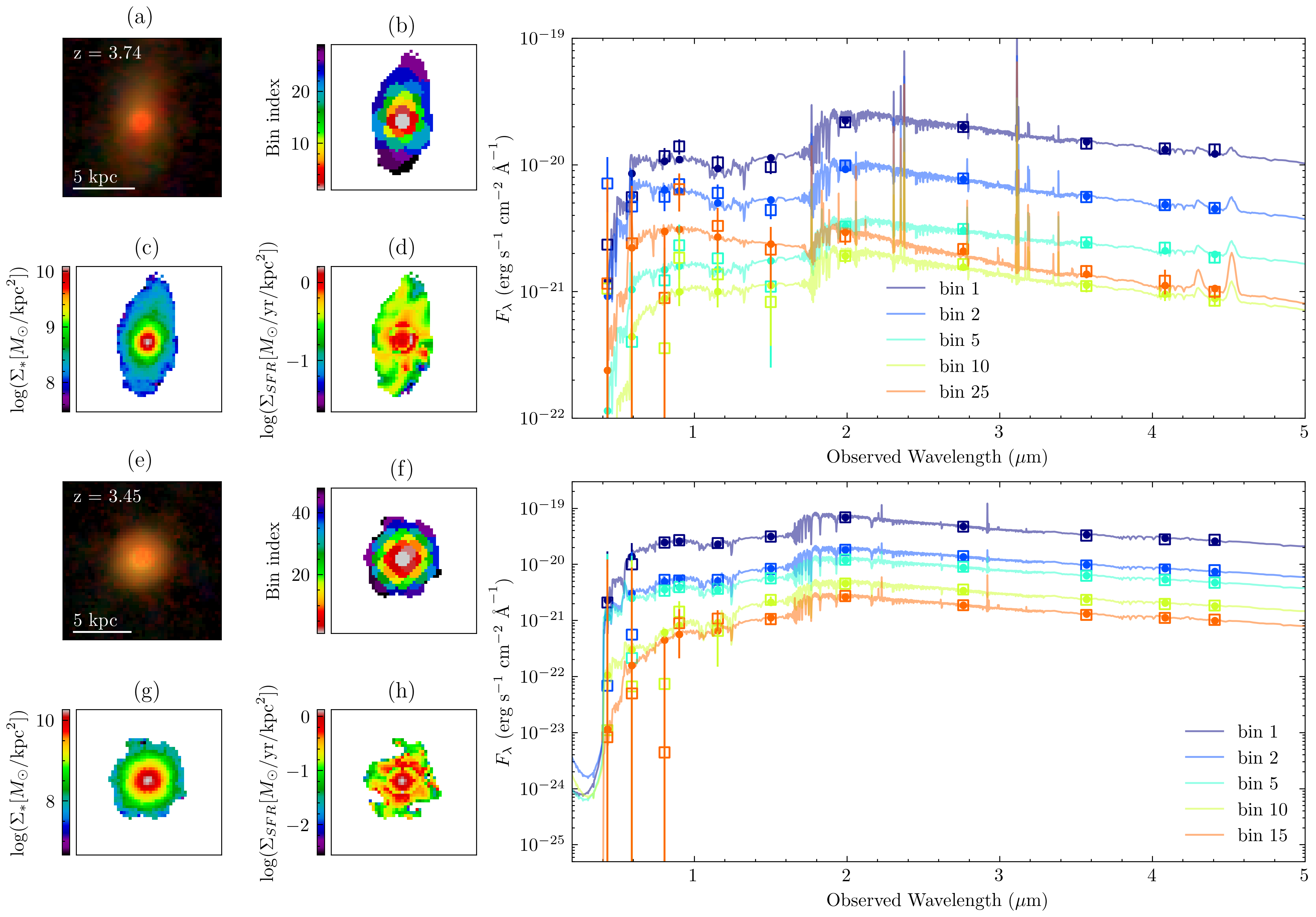}
    \caption{Examples of RGB images (F150W+F200W+F444W) (panels a and e), pixel-binning maps (panels b and f), stellar mass surface density maps (panels c and g), SFR surface density maps (panels d and h), and fitted SED in several bins of primer-cosmos-east-grizli-v7.0 \#45144 (top) and primer-uds-south-grizli-v7.2 \#16789 (bottom) derived with \pixedfit. In the fitted SEDs, open squares represent the observed fluxes, while solid dots indicate the fluxes predicted by the best-fit model.}
    \label{fig:pixedfit_fitted}
\end{figure*}

\begin{table*}
    \centering
    \caption{Assumed priors in the SED modelling}
    \label{tab:fitting}
    \begin{tabular}{l l l l}
    \hline
        Parameter & Description & Prior & Sampling \\
        \hline
        $\log(M_*)$ & Stellar mass & Uniform: min $=\log(s_{best})-2$, max $=\log(s_{best})+2$ & Logarithmic \\
        
        $\log (Z_*/Z_\odot)$ & Stellar metallicity & Uniform: min $=-0.2$, max $=0.2$ & Logarithmic \\
        
        $\log(t)$ & Time since star formation started & Uniform: min $=0.1$, max = age of the universe & Logarithmic \\
         & & at the closest lowest integer redshift & \\

        $\log(\tau)$ & Star-formation  decaying timescale & Uniform: min $=-1$, max $=1.5$ & Logarithmic \\

        $\hat{\tau}_1$ & Birth cloud dust attenuation & Uniform: min $=0.0$, max $=0.6$ & Linear \\
        & in \cite{2000ApJ...539..718C} & & \\

        $\hat{\tau}_2$ & Diffuse ISM dust attenuation & Uniform: min $=0.0$, max $=0.6$ & Linear \\
        & in \cite{2000ApJ...539..718C} & & \\

        $n$ & Power-law index in & Uniform: min $=-2.2$, max $=0.4$ & Linear \\
        & \cite{2000ApJ...539..718C} & & \\
        
        \hline
    \end{tabular}
    \raggedright
    NOTE: $s_{best}$ is the normalization of model SED derived from the initial fitting with the $\chi^2$ minimization method (see Section 4.2.1 in \citealt{2021ApJS..254...15A}).
\end{table*}

\subsection{Constructing Photometric Data Cube}

We utilize \pixedfit\ to generate photometric data cubes from the cutout images across multiple filters. First, we define the region of interest for each galaxy by creating segmentation maps in all filters using \texttt{SEP} with consistent parameters: a detection threshold (\texttt{thresh} = 2.0), which selects pixels $\geq2\sigma$ above the background; the number of thresholds for deblending (\texttt{deblend\_nthresh} = 40), controlling how finely overlapping sources are separated; and the minimum contrast ratio (\texttt{deblend\_cont} = 0.01), allowing components with $\geq1\%$ of the parent flux to be deblended as separate sources.

Segmentation maps can vary significantly between filters due to differences in brightness and signal-to-noise ratio (S/N) at different wavelengths. A simple combination of segmentation maps from all filters can lead to suboptimal results, either producing regions that are too small or too large, sometimes even including nearby sources. To mitigate this issue, we found that the most effective approach was to use the segmentation map from 3 filters closest to the rest-frame visual wavelength ($\sim$500 nm). This choice is particularly suitable for quiescent galaxies, as they are expected to be brightest in this wavelength range. {The final segmentation map is created by taking the combination of these three maps, meaning a pixel is included if it appears in at least one of them.}

After defining the galaxy's region, \pixedfit\ calculate the flux for each pixel within that area. The pixel values are converted to flux densities in units of erg s\(^{-1}\) cm\(^{-2}\) \AA\(^{-1}\) using the PHOTFLAM keyword from the \texttt{GRIZLI} imaging data product headers. This process is performed across all filters, generating a data cube, which is then stored in a FITS file. To ensure robust SED fitting, we exclude galaxies that are fully covered in fewer than six filters. We also perform visual inspection of the galaxy images and remove objects that are cropped or contaminated by light from nearby sources, such as bright stars or neighboring galaxies. After applying these criteria, our final sample consists of 264 galaxies.

\subsection{Pixel Binning}
\label{sec:binning}

Next, we perform pixel binning, which is essential because individual pixel SEDs often have low S/N, making SED fitting unreliable if it is performed on the pixel SEDs. Using \pixedfit\, we group neighboring pixels with similar SED shapes to achieve a target S/N threshold. The core principle of this method is to merge adjacent pixels with similar SED shapes until the specified S/N threshold is reached.

A key advantage of \pixedfit\ is that it allows pixel binning not only based on S/N but also on the similarity of their SEDs. This approach ensures that the resulting binned regions maintain the essential spatial information while increasing the S/N. For a more detailed explanation of the binning technique, see \cite{2017MNRAS.469.2806A} and \cite{2021ApJS..254...15A}.

For the binning process, we apply the following criteria. We set a minimum bin diameter of 5 pixels, which is comparable to the largest PSF FWHM in our dataset. To ensure SED similarity within each bin, we impose a reduced \(\chi^2\) limit of 5. {We set a signal-to-noise (S/N) threshold of 5 for the NIRCam filters and 0 for all ACS filters, reflecting the generally lower S/N in HST images compared to JWST. Additionally, for all filters probing rest-frame wavelengths shorter than the 4000\AA\ break, we assign an S/N of 0, as quiescent galaxies are expected to be very faint in this regime}. {This process results in 91,874 bins. Figure \ref{fig:bin_per_galaxy} shows the number of bins per galaxy, with blue points indicating median values in each redshift bin. The decline at higher redshift likely reflects smaller galaxy sizes, lower stellar masses, and cosmic surface brightness dimming.}.

Examples of the binned images produced by \pixedfit\ are shown in Figure \ref{fig:pixedfit_fitted}. Specifically, panel b shows the galaxy primer-{cosmos-east-grizli-v7.0 \#45144 with 29 spatial bins}, while panel d displays primer-{uds-south-grizli-v7.2 \#16789 with 48 bins}. Each color represents a different bin, ranging from reddish tones in the central regions to purplish hues in the outskirts.

\subsection{Spatially Resolved SED Fitting} \label{sec:SED_fitting}

The next step is to fit the SED of each bin in the binned data cube. This is performed using the SED-fitting module in \pixedfit, which employs a fully Bayesian approach.

\pixedfit\ utilizes the Flexible Stellar Population Synthesis code (\texttt{FSPS}; \citealt{2009ApJ...699..486C}, \citealt{2010ascl.soft10043C}), which includes nebular emission modeling \citep{2017ApJ...840...44B} based on the \texttt{CLOUDY} code \citep{1998PASP..110..761F, 2013RMxAA..49..137F}. In our analysis, we adopt the initial mass function (IMF) from \cite{2003PASP..115..763C}, Padova isochrones \citep{2000A&AS..141..371G, 2007ASPC..374...33M, 2008A&A...482..883M}, and the MILES stellar spectral library \citep{2006MNRAS.371..703S, 2011A&A...532A..95F}. We model the star formation history in the form of the delayed-tau, parameterized as
\begin{equation}
    SFR(t)\propto t e^{-t/\tau}.
\end{equation}
Dust attenuation is modeled using the two-component attenuation law from \cite{2000ApJ...539..718C}. Since our dataset does not include mid-infrared or far-infrared photometry, we switch off the dust emission and AGN dusty torus emission models. Additionally, we fix the ionization parameter to be $U=0.01$ in the nebular emission modeling.

A summary of the assumed parameters and priors is provided in Table \ref{tab:fitting}. We apply the random dense sampling of parameter space (RDSPS) method, which provides faster computation than the Markov Chain Monte Carlo (MCMC) approach while still yielding robust parameter inferences \citep{2021ApJS..254...15A}. SED fitting is performed on each spatial bin in the galaxies. 

Once the fitting has been done to all spatial bins in a galaxy, maps of stellar population properties can then be constructed. We only focus on the maps of stellar mass and SFR in this work. A unique feature of \pixedfit\ is its ability to recover the original spatial resolution of the input images after performing SED fitting on the binned data. Specifically, the total stellar mass and SFR derived for each bin are redistributed back to individual pixels within that bin. This redistribution is guided by the rest-frame red-band flux for stellar mass (F444W) and blue-band flux for SFR (the band with rest-wavelength closest to 2000 \AA), preserving the spatial structure traced by the observed photometry. As a result, the final stellar mass and SFR maps retain the same sampling and resolution as the original photometric images with largest PSF size.

Figure \ref{fig:pixedfit_fitted} presents examples of the resulting parameter maps that include stellar mass surface density ($\Sigma_*$) and SFR surface density ($\Sigma_{\mathrm{SFR}}$), along with the fitted SEDs. Panels c and d correspond to primer-cosmos-east-grizli-v7.0 \#45144, while panels g and h show results for primer-uds-south-grizli-v7.2 \#16789. In the fitted SEDs, open squares represent the observed fluxes, while solid circles indicate the fluxes predicted by the best-fit models.

To ensure our sample consists solely of quiescent galaxies, based on the estimated SFR and stellar mass from \pixedfit\, we exclude objects with total SFR exceeding the star formation main sequence \citep{2014ApJS..214...15S} at their given stellar mass and redshift. Our final sample consists of {256 galaxies (86,696 bins)}, with 101 of them having confirmed spectroscopic redshifts.

\section{Results} \label{sec:results}

\begin{figure*}
    \centering
    \includegraphics[width=0.9\linewidth]{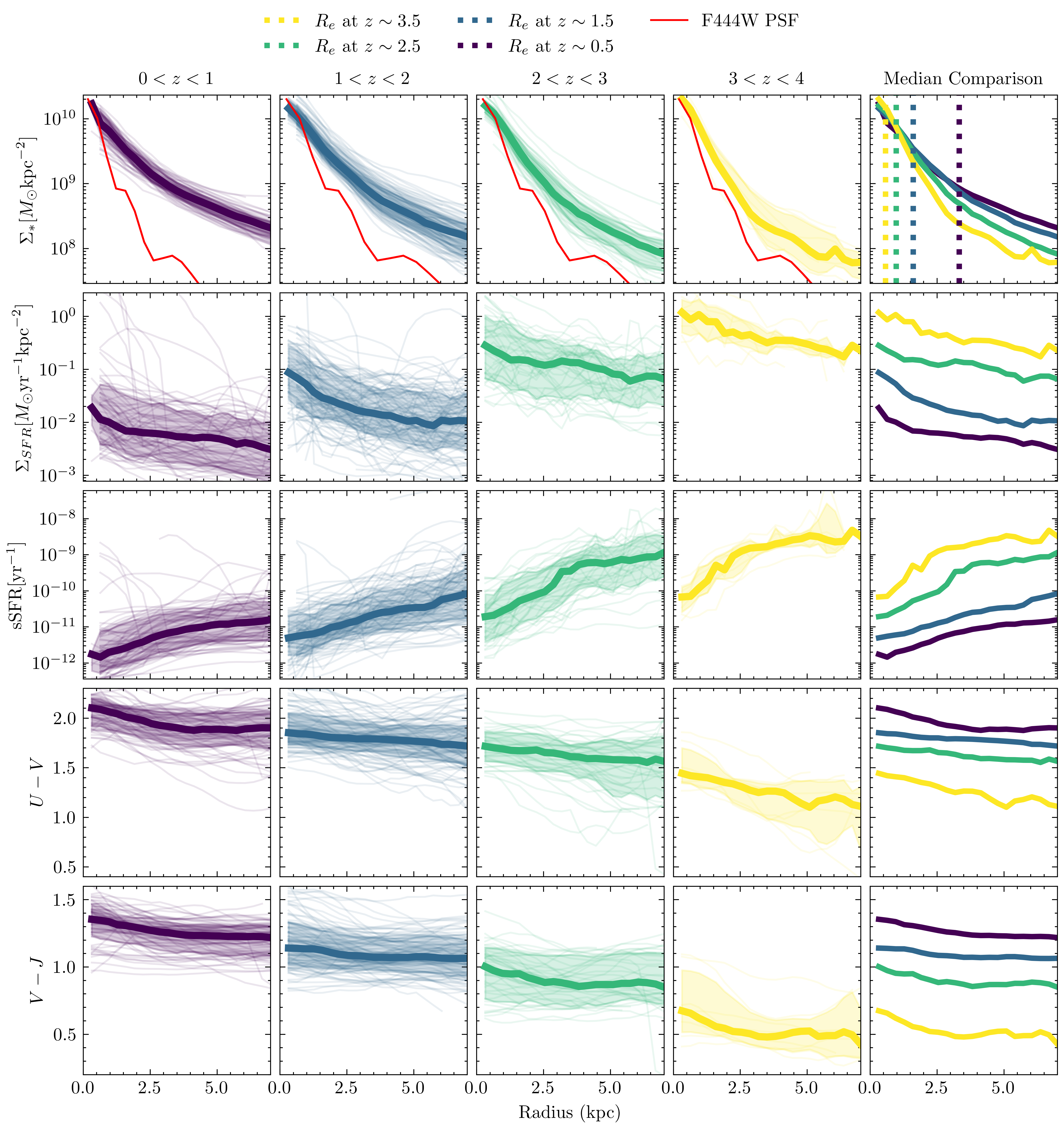}
    \caption{Radial profiles of stellar mass surface density ($\Sigma_*$, top row), SFR surface density ($\Sigma_{SFR}$, second row), specific SFR (sSFR, third row), $U-V$ (fourth row), and $V-J$ colors (fifth row) from \pixedfit, color-coded by redshift. Thick lines represent the median values, while shaded regions indicate the 16th to 84th percentile range. Fifth column presents the median profiles for each redshift bin. The dotted lines in fifth column indicate the half-mass radius (top row). The solid red lines in the uppermost row showing the largest PSF profile, obtained from F444W band.}
    \label{fig:profiles}
\end{figure*}

\begin{figure*}
    \centering
    \includegraphics[width=\linewidth]{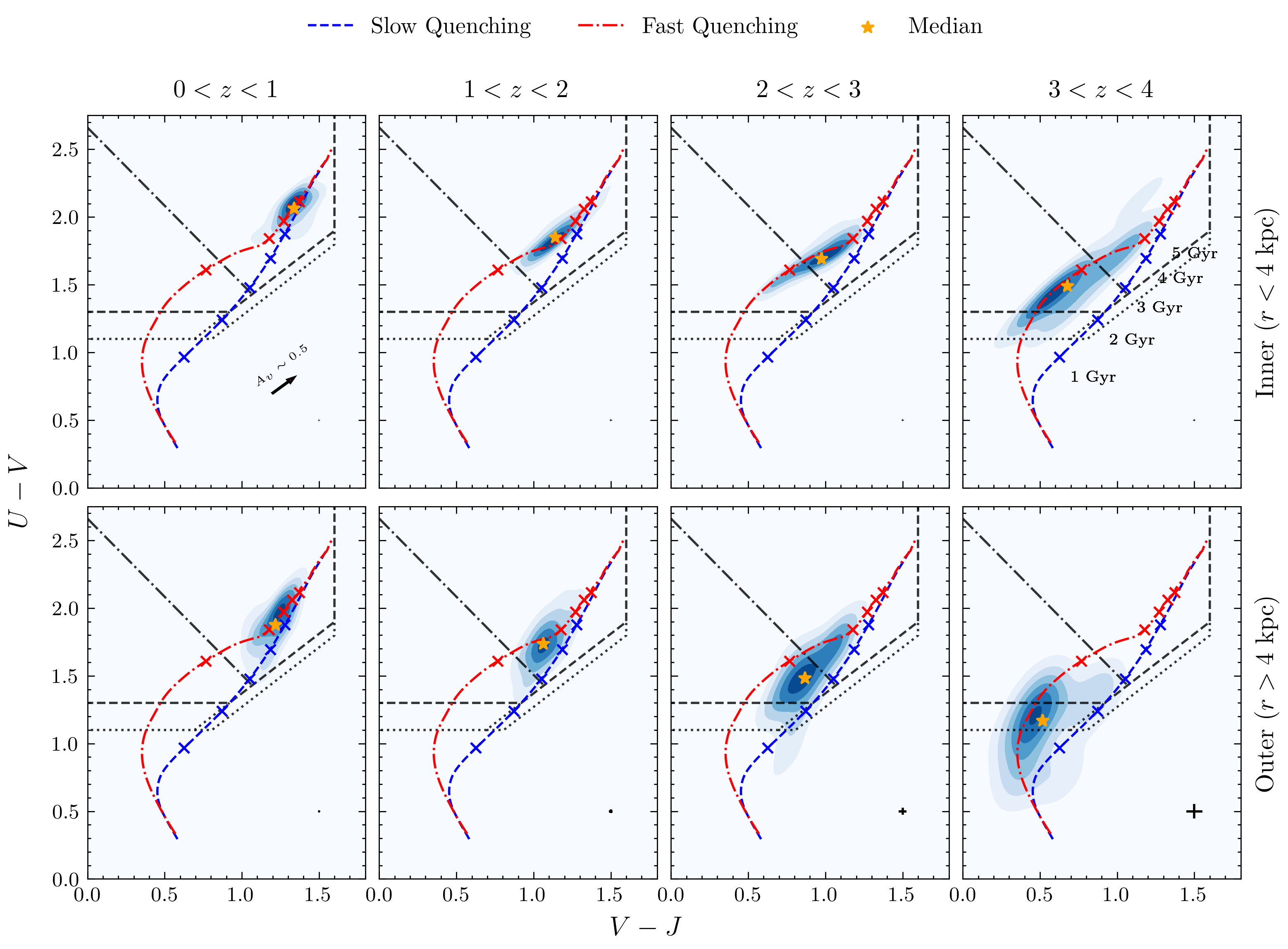}
    \caption{Evolution of the sample galaxies in the \textit{UVJ} diagram across different redshift bins (\(0 < z < 4\)). Each row represents different radial regions: inner (\(r < 4\) kpc, top row), and outer (\(r > 4\) kpc, bottom row). Each panel shows the density of galaxies in $U-V$ and $V-J$ color space for the corresponding radial region, visualized using filled contours. The {orange star} symbol in each panel indicates the median value. The dashed, dash-dotted, and dotted black lines indicate the quiescent region from \cite{2009ApJ...691.1879W}, fully-quenched and recently quenched galaxies separator line from \cite{2017MNRAS.472.1401A}, and the extended region for recently quenched galaxies from \cite{2022A&A...666A.141M}. The evolutionary tracks for slow (blue dashed) and fast (red dashed-dotted) quenching models are overlaid for comparison. Red and blue cross symbols mark the positions of galaxies at ages of 1, 2, 3, 4, and 5 Gyr. Median $U-V$ and $V-J$ color errors, shown at the bottom right of each panel, are estimated from observed fluxes using filters with rest-frame wavelengths closest to the $U$, $V$, and $J$ bands. The black arrow in the top-left panel indicates the dust attenuation vector corresponding to \( A_V \sim 0.5 \).}
    \label{fig:UVJfitted}
\end{figure*}

\begin{table}
    \centering
    \caption{The number of galaxies and the median values of the half-mass radius in each redshift bin are shown. Uncertainties indicate the scatter in the data, estimated through bootstrapping of the medians.}
    \label{tab:rad}
    \begin{tabular}{l l l l}
    \hline
        Redshift Range & \#Galaxies & Half-Mass Radius (kpc) \\
    \hline
        $0 < z < 1$ & 84 & $3.35\pm0.23$ \\
        $1 < z < 2$ & 118 & $1.61\pm0.07$ \\
        $2 < z < 3$ & 45 & $0.99\pm0.07$ \\
        $3 < z < 4$ & 9 & $0.61\pm0.09$ \\
        \hline
    \end{tabular}
\end{table}

\subsection{Radial Stellar Mass and Star Formation Profiles} \label{sec:fitted_prof}

To investigate how the internal stellar mass structure and star formation activity of massive quiescent galaxies evolve with redshift, we analyze the spatially resolved distributions of $\Sigma_*$, $\Sigma_{\rm SFR}$, and specific SFR (sSFR).{ Radial profiles were derived from the parameter maps by averaging pixel values within a series of concentric circular annuli centered on each galaxy, with a fixed bin width of 0.3 kpc}. These profiles offer insight into the buildup of stellar mass and the spatial distribution of star formation across different cosmic epochs.

{Figure \ref{fig:profiles} presents the radial profiles of key parameters for all sample galaxies, grouped into four redshift bins. Thick solid lines indicate the median profiles, with shaded regions showing the 16th–84th percentile range. The rightmost column compares profiles across redshifts. At $z \sim 3\text{–}4$, profiles extend only to $r \sim 7$ kpc, while lower-redshift galaxies reach out to $\sim 20$ kpc. As discussed in Section \ref{sec:binning}, this difference is likely due to a combination of cosmic surface brightness dimming and the intrinsically smaller, fainter nature of high-redshift galaxies. To ensure a fair comparison across bins, we restrict all profiles in the figure to within $r \sim 7$ kpc.}

The stellar mass surface density profiles exhibit steeper gradients at higher redshifts, indicating more centrally concentrated stellar mass distributions. At lower redshifts, the profiles become flatter, reflecting increased stellar mass at larger radii. While the outer (\(r>4\) kpc) regions tend to have higher stellar mass densities in lower redshift compared to higher redshift bins, the inner region (\(r<4\) kpc) maintains a consistent stellar mass density across redshift bins, with a central peak of approximately \(10^{10} M_{\odot} \, \text{kpc}^{-2}\).  

The SFR surface density profiles  exhibit a decline in overall amplitude with decreasing redshift, indicating reduced star formation activity across all radii. In addition, the sSFR radial profiles reveal that at higher redshifts ($z>2$), the outskirts of galaxies have significantly higher sSFR values than the central regions, with differences exceeding 1 dex. As redshift decreases, the overall sSFR declines, and the distinction between the inner and outer regions diminishes.

In the rightmost column of Figure~\ref{fig:profiles}, we show the median half-mass radii (first row), indicated by vertical dashed lines and color-coded by redshift. The half-mass radius is defined as the radius enclosing half of the total stellar mass. We measure these radii by fitting the stellar mass maps using \texttt{statmorph} \citep{2019MNRAS.483.4140R}, which accounts for the effects of the PSF on the stellar mass distribution. This correction is particularly important because, at the highest redshift bin, the stellar mass profiles in Figure~\ref{fig:profiles} approach the PSF profile. The median half-mass radii are summarized in Table~\ref{tab:rad}. We find that the half-mass radii grow with cosmic time, increasing by {a factor of 5.4} from \(z \sim 3.5\) to \(z \sim 0.5\).

\subsection{Color Distributions} \label{sec:UVJ}

In this subsection, we examine how the spatial distribution of colors within galaxies varies with redshift and use this to confirm the trends in spatially resolved star-forming activity observed in the previous subsection. Using the best-fit model spectra from the SED fitting of each bin, we compute the rest-frame \(U\), \(V\), and \(J\) fluxes. The \(U\) and \(V\) magnitudes are defined following \cite{2006AJ....131.1184M}, while the \(J\) magnitude is computed using the 2MASS \textit{J}-band transmission curve \citep{2006AJ....131.1163S}. {To estimate the flux contribution at the pixel level, we distribute the modeled \(U\), \(V\), and \(J\) fluxes of each bin to individual pixels by scaling them according to the observed flux in each pixel at the rest-frame wavelength closest to the target band, normalized by the total observed flux within the bin.}

From the pixel-level estimates, we derive the radial profiles of the $U\!-\!V$ and $V-J$ colors, shown in the fourth and fifth rows of Figure~\ref{fig:profiles}, respectively. As seen in the figure, galaxies at lower redshift are redder in both $U\!-\!V$ and $V-J$ compared to those at higher redshift. The $U\!-\!V$ profiles are steep and centrally peaked at the highest redshift bin, becoming progressively flatter toward lower redshifts. Since the $U\!-\!V$ color is sensitive to the age of the stellar population, this indicates that at $3<z<4$, the outer regions host younger stellar populations compared to the centers, while the age difference becomes less pronounced at lower redshifts. In the $V-J$ radial profiles, we also find a consistent trend of bluer colors toward the outer regions across all redshift bins.

We also calculate the integrated \(U - V\) and \(V - J\) colors within two radial regions of each galaxy: an inner region (\(r < 4\) kpc) and an outer region (\(r > 4\) kpc). The division at 4 kpc is chosen to mitigate contamination from the PSF wings (see first row of Figure \ref{fig:profiles}), which can affect photometric measurements in the outer region. The resulting spatial color distributions are presented in Figure \ref{fig:UVJfitted}. In Figure \ref{fig:UVJfitted}, the top and bottom rows represent the colors of the inner and outer regions of galaxies, respectively. Each column corresponds to a different redshift bin, with the lowest redshift on the left and the highest on the right. The {orange star} symbol in each panel indicates the median value. The blue dashed and red dot-dashed lines indicate the predicted evolutionary tracks of slow and fast quenching processes, modeled using \pixedfit. These tracks were generated by computing rest-frame spectra of galaxies, using the same configurations as the SED fitting described in Section \ref{sec:SED_fitting}, with fixed parameters: \(\tau_1\) and \(\tau_2\) set to 0.4, a dust attenuation slope of \(n=-0.7\), a stellar metallicity of \(\log(Z/Z_\odot)=-0.1\), and star formation timescales of \(\tau=0.1\) Gyr and \(\tau=1\) Gyr for the fast and slow quenching tracks, respectively. The only free parameter was the stellar population age, which was sampled across 100,000 time steps from the onset of star formation. To get the evolution of $UVJ$ colors, we generate a model spectrum at each time step and then calculate the rest-frame $U-V$ and $V-J$ colors. The crosses along these tracks indicate the predicted locations in the \textit{UVJ} diagram at various times since the onset of star formation (1, 2, 3, 4, and 5 Gyr, from bottom to top, respectively). The black arrow in the top-left panel indicates the dust attenuation vector corresponding to \( A_V \sim 0.5 \). 

We include additional lines in this figure alongside the original quiescent box from \cite{2009ApJ...691.1879W} to show variations in galaxy quenching phase. A diagonal dash-dotted line is added within the quiescent box to separate fully quenched galaxies (above the line) from post-starburst or recently quenched galaxies (below the line), following \cite{2014MNRAS.440.1880W} and \cite{2017MNRAS.472.1401A}. Additionally, the solid horizontal boundary is extended to \(U-V=1.1\), and the original diagonal line is shifted by \(\Delta(U-V)=0.1\) (dotted lines) {to show a region that captures recently quenched galaxies just outside the original quiescent box}, as performed in \cite{2022A&A...666A.141M}.

In the lower right corner of each panel, we show the median errors of the $U-V$ and $V-J$ colors. These errors are derived from the observed fluxes, as \pixedfit\ does not provide uncertainties for the fitted spectra. To estimate the errors, we use the observed filters whose rest-frame wavelengths are closest to the $U$, $V$, and $J$ bands. We note that most of the uncertainties are very small and therefore not visible in the plot.

The color distributions of galaxies exhibit a distinct inside-out quenching pattern, with systematic differences between their inner and outer regions. In the highest redshift bin (\(3<z<4\)), most inner regions fall within the quiescent box, whereas almost all outer regions remain outside it. At lower redshifts, the fraction of outer regions inside the quiescent box increases, and by the lowest redshift bin (\(0<z<1\)), nearly all regions in our sample appear quiescent. These trends align with the observed sSFR profiles in Figure \ref{fig:profiles}, where the inner regions show lower sSFR compared to the outer parts, and the profiles gradually flatten over time.

Moreover, the evolutionary tracks of these regions suggest distinct quenching mechanisms. The inner regions align nicely with the fast quenching model, in which star formation is rapidly suppressed. In contrast, {the outer regions exhibit greater scatter and some galaxies tend to align more closely with the slow quenching model}, where star formation declines gradually over an extended period.

{However, since galaxies at lower redshifts intrinsically tend to have higher stellar masses than their higher-redshift counterparts, we note that the redshift-dependent trends discussed in this and the previous section may also be influenced by these stellar mass differences.}

\begin{figure*}
    \centering
    \includegraphics[width=\linewidth]{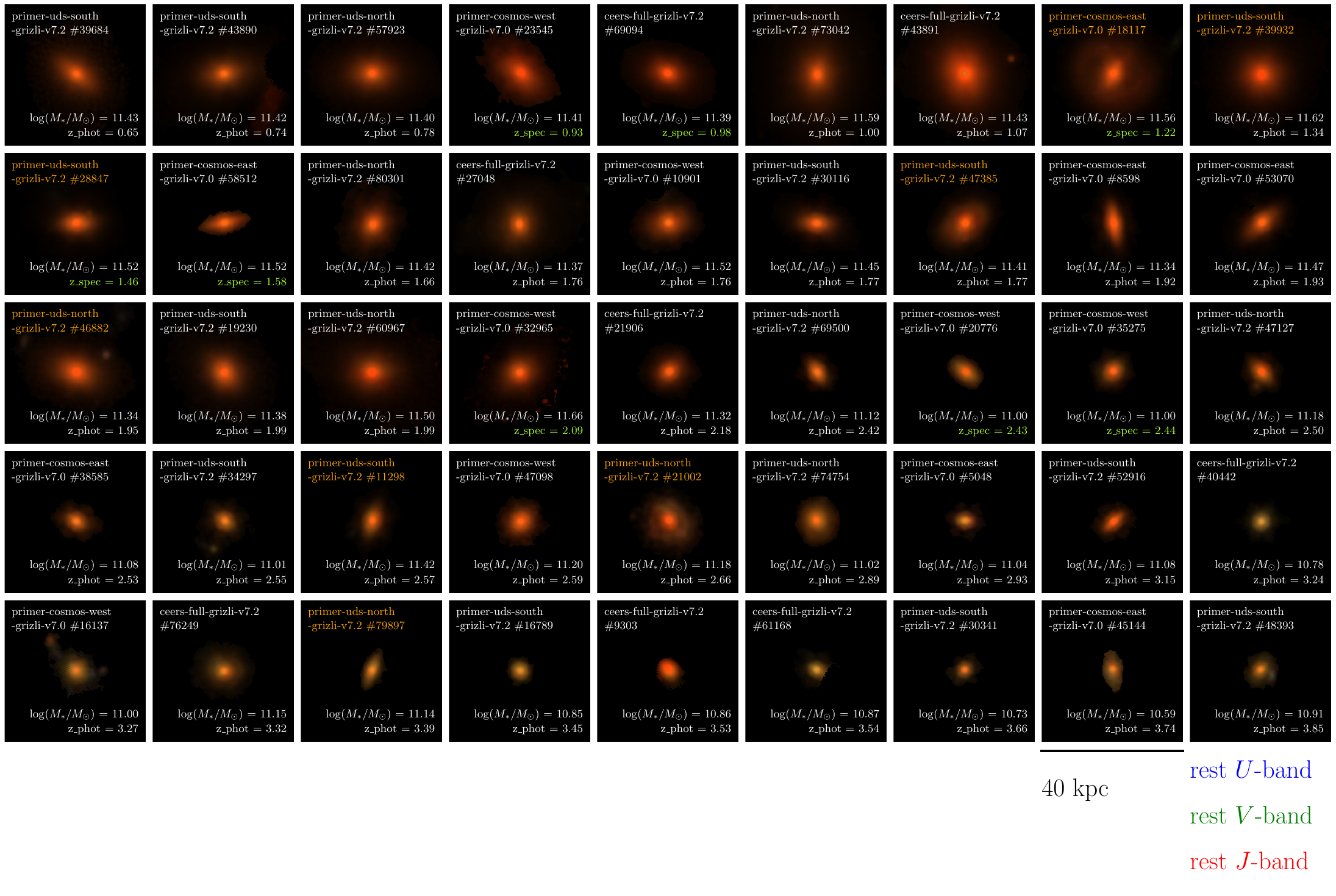}
    \caption{Color-composite images of the most massive galaxies in each redshift bin (see Section \ref{sec:progen}). These images are constructed using the estimated rest-frame \(U\), \(V\), and \(J\) bands fluxes derived from \texttt{piXedfit}, mapped to the blue, green, and red components, respectively. The method for estimating the rest-frame fluxes for each pixel is described in Section \ref{sec:UVJ}. Each panel is scaled to a physical size of $\sim40$ kpc. The redshift and estimated total stellar mass of each galaxy are indicated in the corresponding panel; spectroscopic redshifts are shown in green. Galaxies hosting X-ray detected sources are labeled in orange.}
    \label{fig:cutouts}
\end{figure*}

\begin{figure}
    \centering
    \includegraphics[width=\linewidth]{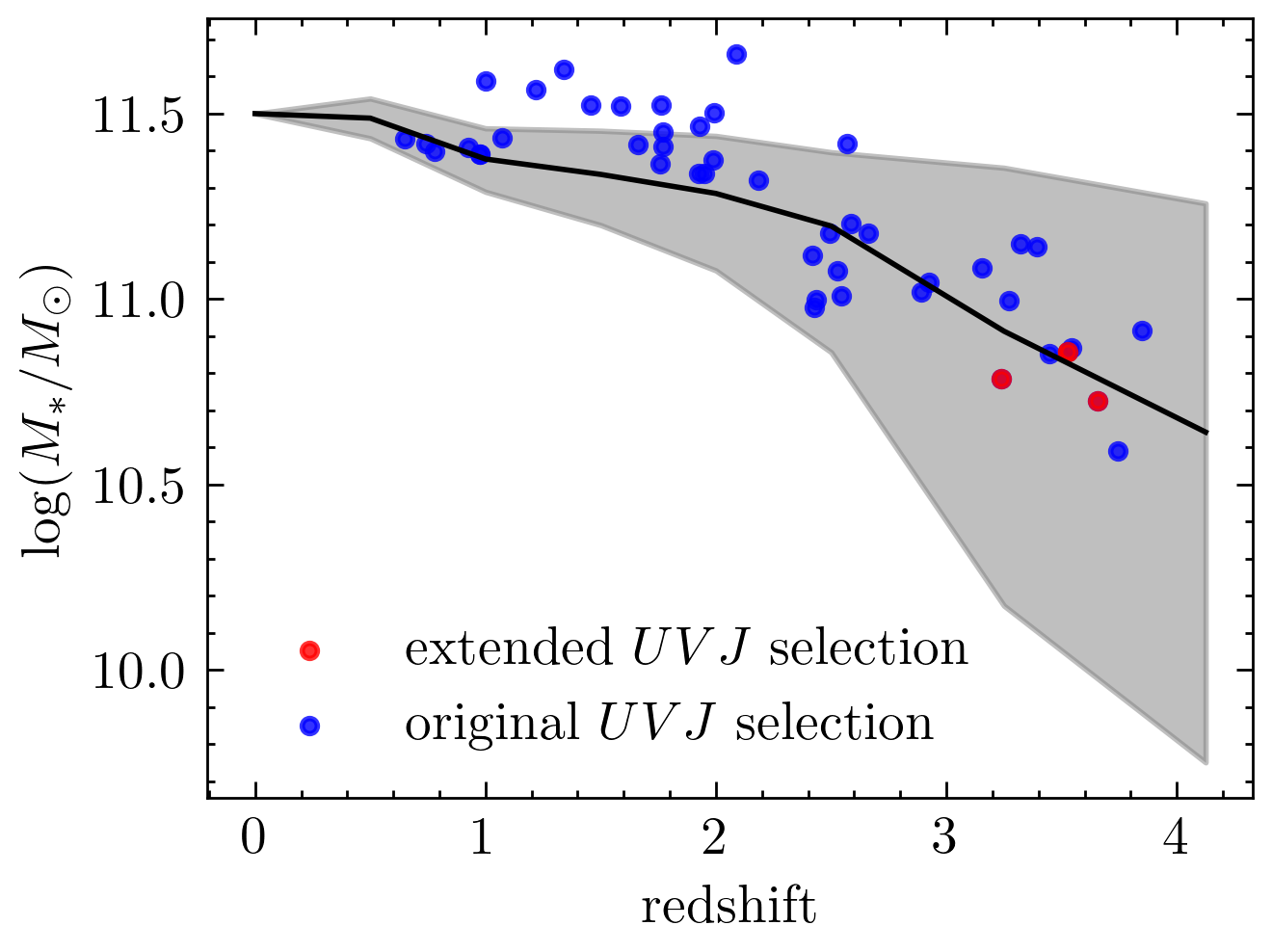}
    \caption{Predicted (solid black line) and observed stellar mass of our sample in Section \ref{sec:progen}. Red point represents galaxy selected using the extended $UVJ$ selection, while blue points represent those selected using the original $UVJ$ criteria. The predicted stellar mass evolutions are calculated with the abundance matching method using the \texttt{nd-redshift} code \citep{Behroozi2013} and the stellar mass functions from \citet{Baldry2012}, \citet{McLeod2021}, and \citet{Weibel2024}.}
    \label{fig:massevol_comp}
\end{figure}

\subsection{Progenitors-Descendants Evolution}\label{sec:progen}

In order to examine the structural evolution of massive quiescent galaxies quantitatively with defining ancestors and descendants, we further select the {11-12} most massive galaxies in each redshift bin, with the number of galaxies adjusted according to the comoving volume of each bin. This approach assumes that the most massive galaxy in our highest redshift bin remains the most massive descendant in each subsequent lower redshift bin \citep[see, e.g.,][]{2010ApJ...709.1018V, 2011MNRAS.412.1123P}. Table \ref{tab:comov} lists the redshift ranges, corresponding comoving volumes, and the number of selected galaxies. 

{However, our progenitor–descendant sample at lower redshifts may be contaminated by massive galaxies that have only recently quenched and are not true descendants of the quiescent galaxies in the highest redshift bin. These newcomers may have stellar masses comparable to or even higher than those of the evolved quiescent galaxies. If this is the case, our analysis could be biased toward more massive, compact, and centrally dense systems \citep[e.g.,][]{2017MNRAS.472.1401A}.
}

In this analysis, we extend our sample at \(z > 3\) by adjusting the \(U - V\) threshold from 1.3 to 1.1 and shifting the diagonal boundary downward by 0.1 to include recently quenched galaxies, following \cite{2022A&A...666A.141M}. This was done because at \(z \gtrapprox 3\), recently quenched massive quiescent galaxies with short quenching timescales may lie just below the traditional quiescent box rather than inside it \citep[e.g.,][]{2017MNRAS.472.1401A, 2023ApJ...947...20V, 2023MNRAS.518.5953L}. This adjustment adds {three} additional galaxies at \(z > 3\).

Images of the galaxy sample are shown in Figure \ref{fig:cutouts}. These color-composite images are constructed using the estimated rest-frame \(U\), \(V\), and \(J\) bands fluxes derived from \texttt{piXedfit}, mapped to the blue, green, and red components, respectively. The method used to estimate the rest-frame \(U\), \(V\), and \(J\) fluxes for each pixel within the galaxy images is described in Section \ref{sec:UVJ}. Each panel in the figure is scaled to a physical size of $40$ kpc. The redshift and estimated total stellar mass of each galaxy are indicated in the corresponding panel, with spectroscopic redshifts shown in green. In total, {8/45 of these galaxies} have confirmed spectroscopic redshifts. Galaxies hosting X-ray detected sources are labeled in orange ({8/45 galaxies}).

We present the stellar masses of the selected galaxies in Figure~\ref{fig:massevol_comp}. Red points represent galaxies selected using the extended $UVJ$ selection, while blue points represent those selected using the original $UVJ$ criteria. To validate the progenitors and descendants correspondence of the sample, we compare it with predictions from the abundance matching technique, which traces the stellar mass evolution of progenitor galaxies.

Specifically, we estimate the expected stellar mass evolution of galaxies that reach a present-day mass of $10^{11.5}\,M_{\odot}$. We first compute the co-moving number density of such galaxies at $z \sim 0$ using the stellar mass function from \citet{Baldry2012}, yielding a value of $10^{-5.36} \rm{Mpc}^{-3}$. To trace this number density back to higher redshifts, we use the \texttt{nd-redshift} code from \citet{Behroozi2013}, which determines the median progenitor number density at a certain redshift given an initial number density at a higher redshift, based on peak halo mass functions. This method helps minimize biases from scatter in stellar mass and luminosity. Since \texttt{nd-redshift} does not directly associate stellar masses with number densities, we assign stellar masses at each redshift using stellar mass functions from \citet{McLeod2021} for $0.25 < z < 3.75$ and from \citet{Weibel2024} for $3.75 < z < 9.50$. The resulting median and uncertainty bounds of the stellar mass evolution are shown as a gray line and shaded region in the figure. Most of our sample lies within this predicted range, supporting a possible evolutionary link between our selected galaxies and the expected progenitors of massive local galaxies.

\begin{figure}
    \centering
    \includegraphics[width=\linewidth]{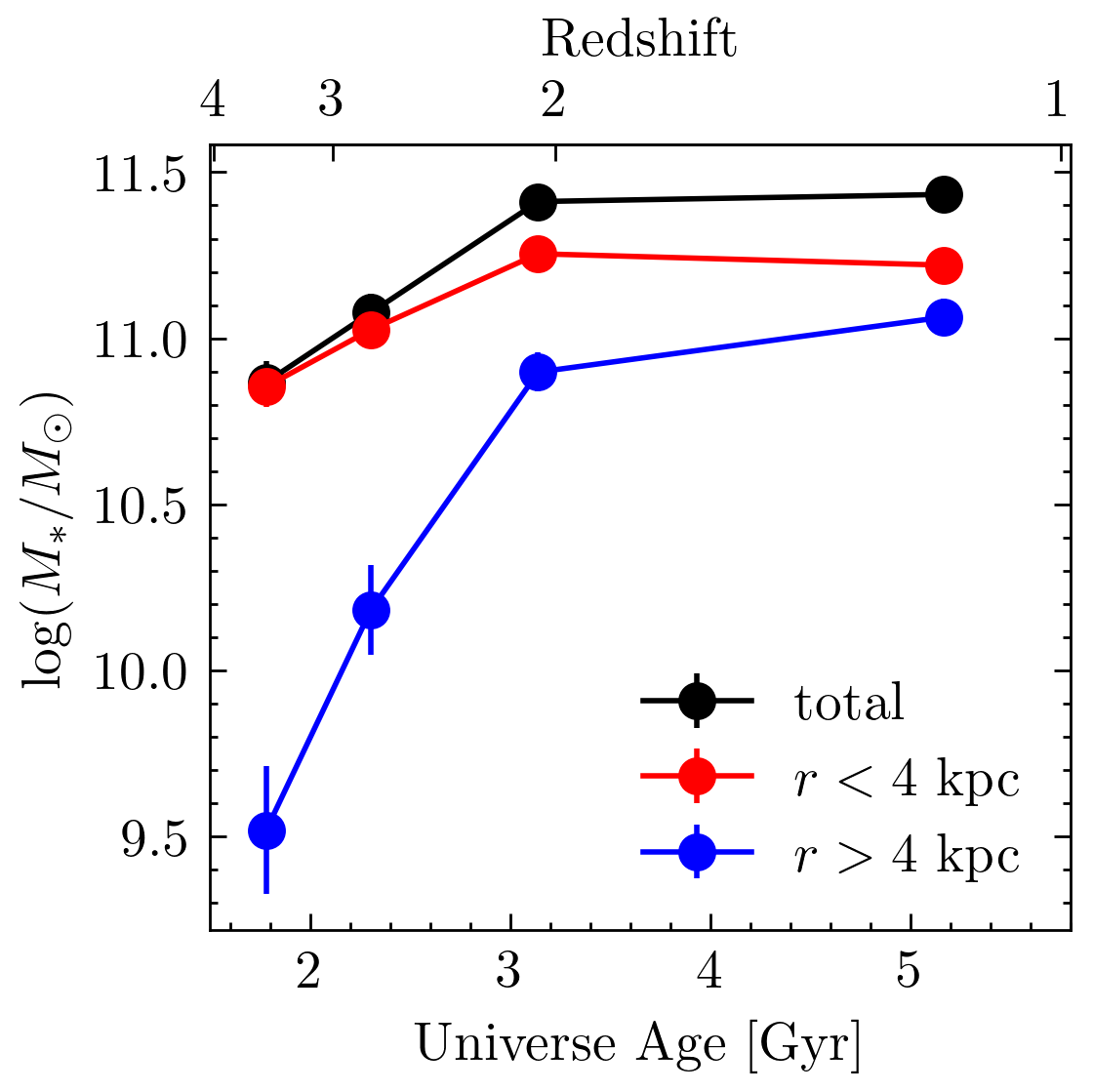}
    \caption{Evolution of stellar mass in massive quiescent galaxies. The figure displays the median stellar mass for the inner (red), outer (blue), and total (black) regions across different redshift bins. Error bars indicate the scatter in the data, estimated through bootstrapping.}
    \label{fig:mass_evol_progen}
\end{figure}

We then calculate the median of stellar mass contained in the inner and outer regions of sample galaxies in each redshift bin. Figure \ref{fig:mass_evol_progen} presents the median stellar mass values for these regions. The red, blue, and black symbols represent the inner, outer, and total stellar masses, respectively. Error bars indicate the scatter in the data, estimated through bootstrapping. {We note that most of the scatter values are smaller than the typical uncertainty from SED fitting ($\sim0.16$) dex.}

In Figure \ref{fig:mass_evol_progen}, the mass of the inner region shows only modest growth over time, whereas the outer region mass increases substantially. This suggests that the overall size evolution of massive quiescent galaxies is predominantly driven by mass accumulation in their outskirts. This trend is consistent with the observed evolution of mass profiles across our entire sample, as shown in Figure \ref{fig:profiles}, where the outer regions of the mass profiles exhibit significant growth and flatten over time, while the inner regions remain relatively unchanged. 

\begin{figure*}
    \centering
    \includegraphics[width=0.7\linewidth]{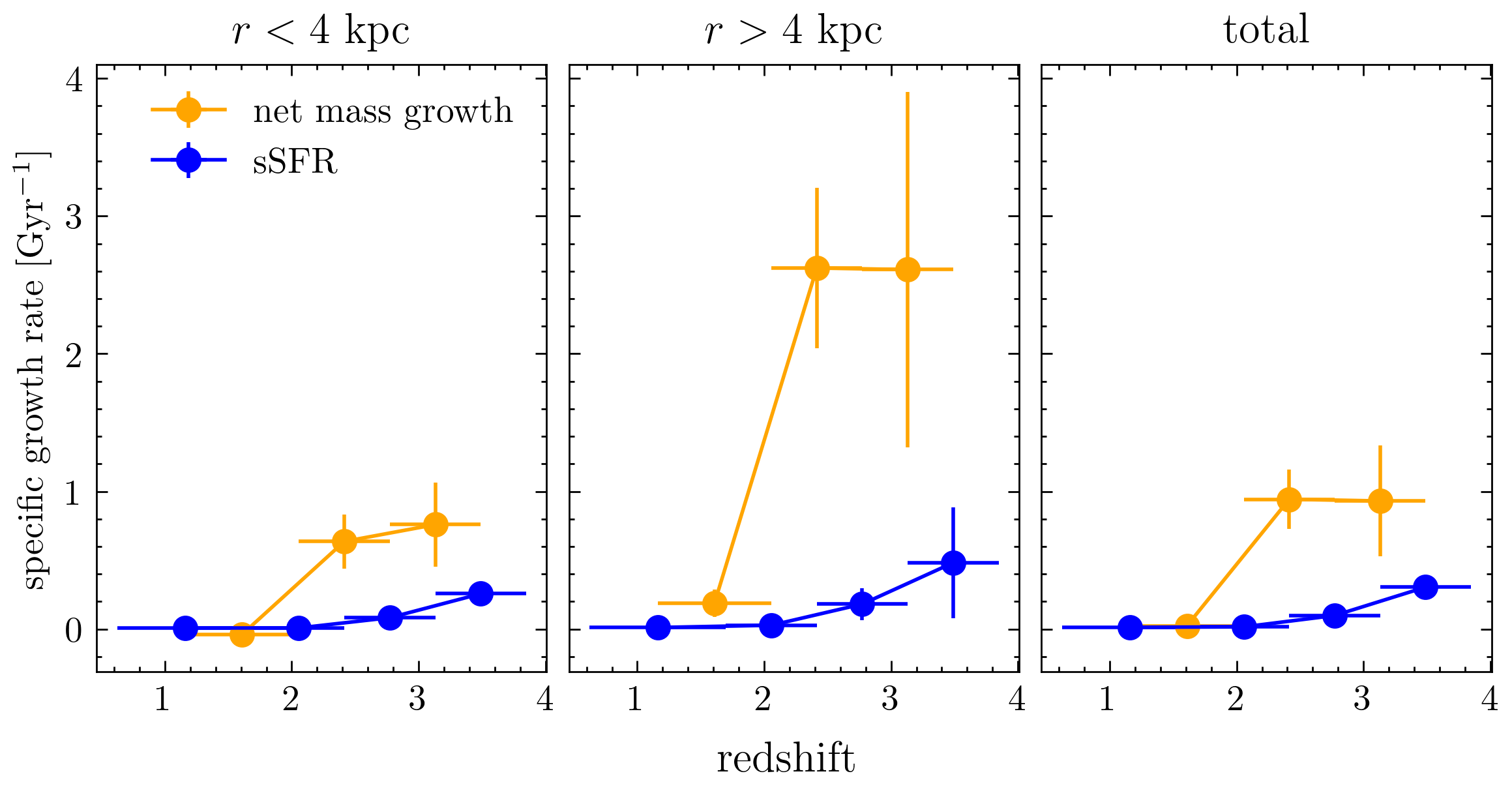}
    \caption{Comparison of required specific mass growth rates (orange) with median of estimated sSFR (blue) for inner (leftmost column), outer (middle column), and total regions (rightmost column). Error bars indicate the scatter in the data, estimated through bootstrapping.}
    \label{fig:SFR_cum_mass}
\end{figure*}

Based on the stellar mass evolution shown in Figure \ref{fig:mass_evol_progen}, {we estimate the required specific mass growth rates by dividing the stellar mass increase between each redshift bin by the corresponding time interval, and normalizing by the average stellar mass between two redshift bins. This quantity can be interpreted as the inverse of the stellar mass doubling time}. In Figure \ref{fig:SFR_cum_mass}, these required specific mass growth rates are depicted in orange, overlaid with the observed sSFR shown in blue. The panels are arranged from left to right to represent the inner, outer, and total regions. Error bars indicate the scatter in the data, estimated through bootstrapping. 

{The specific mass growth rates in the outer regions are significantly higher than those in the inner regions, indicating that stellar mass in the outskirts builds up on much shorter timescales.} Furthermore, these net specific growth rates show that galaxies in all regions experience more rapid stellar mass growth at $z>2$, while below this threshold, {the growth slows considerably. This suggests a transition in the dominant processes governing stellar mass assembly around $z\sim2-3$.}

From the predicted SFR, we estimate the stellar mass formed through in-situ star formation by integrating the SFR over time. Our results show that by $z \sim 1$, in-situ star formation accounts for only {50\%, 8\%, and 36\% }of the stellar mass in the inner, outer, and total regions, respectively. This reveals that in-situ star formation alone is insufficient to explain the total stellar mass growth, especially in the outer regions. This discrepancy indicates that additional processes, most notably mergers, play a crucial role in supplementing stellar mass buildup.

\begin{table*}
    \centering
    \caption{Redshift bin ranges, corresponding comoving volumes, the number of selected galaxies, stellar mass, core surface density, and half-mass radius of the most massive quiescent galaxies. The errors are showing scatter in the data, determined with bootstrapping method.}
    \label{tab:comov}
    \begin{tabular}{l l l l l l}
    \hline
       Redshift Range & Comoving Volume & \#Galaxies & log Stellar Mass & log Core (1 kpc) Surface Density & Half-Mass Radius\\
       & ($\times10^6$ Mpc$^3$) &  & ($M_\odot$) & ($M_\odot\rm{kpc}^{-2}$) & (kpc)\\
       \hline
       0.62 - 1.70 & 1.57 & 12 & $11.43\pm0.04$ & $10.17\pm0.03$ & $3.07\pm0.44$\\
       1.70 - 2.41 & 1.41 & 11 & $11.41\pm0.04$ & $10.21\pm0.05$ & $1.96\pm0.17$\\
       2.41 - 3.13 & 1.43 & 11 & $11.08\pm0.05$ & $10.10\pm0.03$ & $1.02\pm0.13$\\
       3.13 - 3.85 & 1.36 & 11 & $10.87\pm0.07$ & $10.09\pm0.06$ & $0.62\pm0.09$\\
       \hline
    \end{tabular}
\end{table*}

\section{Discussion} \label{sec:discussion}

\subsection{Structural Evolution of Massive Quiescent Galaxies}

Our analysis indicates that the structural evolution of massive quiescent galaxies is primarily driven by mass growth in their outskirts rather than in their central regions. From Figure \ref{fig:mass_evol_progen}, at high redshifts (\(z \sim 3.5\)) the inner regions had already formed approximately {43\% of their stellar mass at $z\sim1$, with only modest growth (\(\sim0.4\) dex) occurring over time. Meanwhile, at $z\sim3.5$, the outer regions contained just \(\sim3\%\)} of stellar mass in the outer region formed by \(z \sim 1\), suggesting that the compact central region formed early, prior to the development of the extended envelopes. The central region formation can be achieved by a gas compaction event, which is a process of gas falling toward the galactic center that leads to a nuclear starburst. This process can be caused by various mechanisms, such as mergers, counter-rotating streams, or low-angular-momentum recycled gas \citep[e.g.,][]{2014MNRAS.438.1870D, 2015MNRAS.450.2327Z, 2016MNRAS.458..242T}. This process led to a high central mass concentration, making the galaxy appear more compact at early times \citep[e.g.,][]{2017MNRAS.472.1401A, 2023ApJ...945..117A}.  

\begin{figure}
    \centering
    \includegraphics[width=\linewidth]{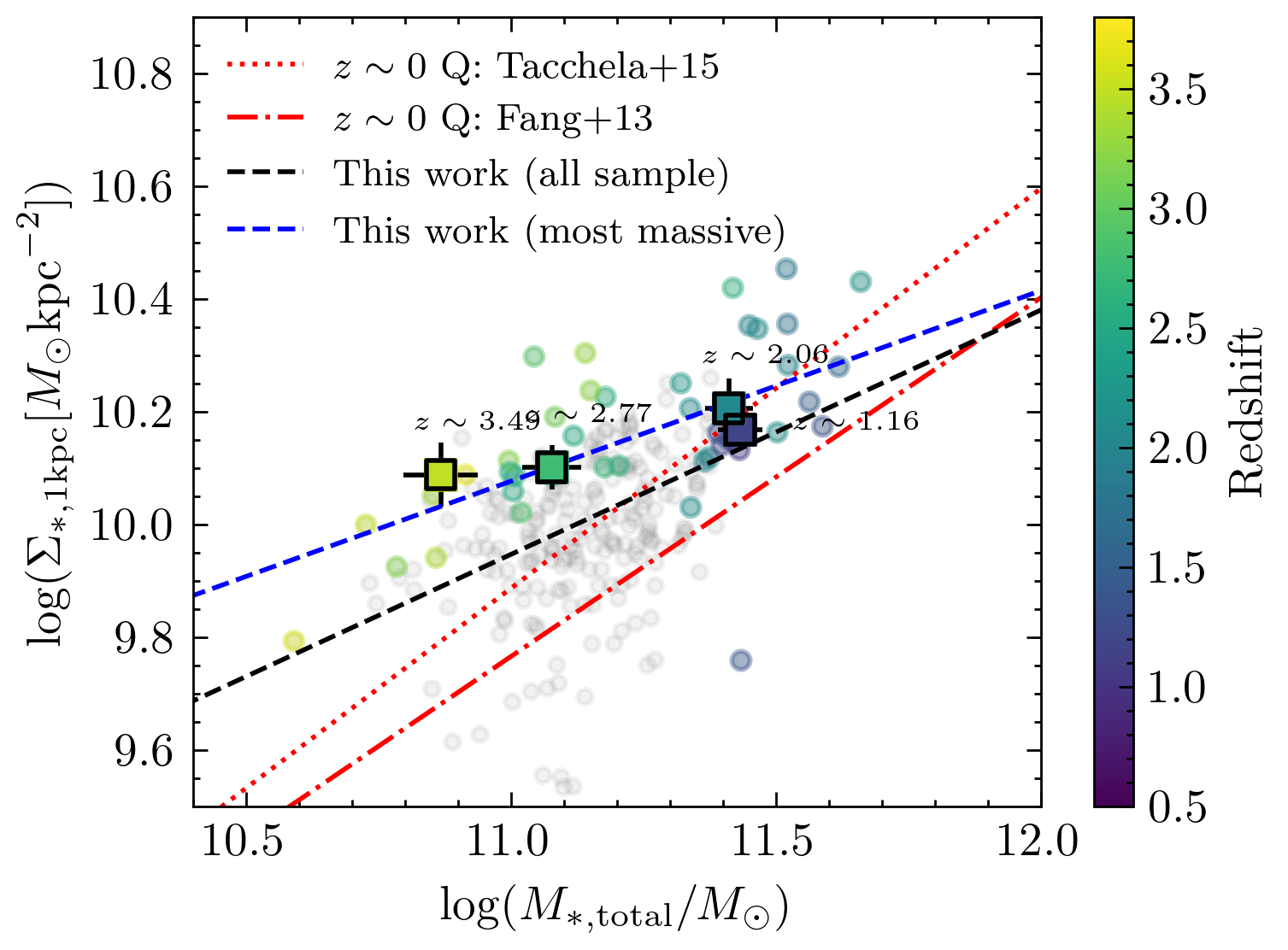}
    \caption{The relationship between core stellar mass surface density (\(\Sigma_{*,1kpc}\)) and total stellar mass for our sample. Colored and grey points represent our progenitor-descendant sample (Section \ref{sec:progen}) and the full sample, respectively. The blue, black, and red lines indicate fitted relations for the progenitor-descendant sample, the full sample, and empirical relations from previous studies (\citealt{2015Sci...348..314T, 2013ApJ...776...63F}).}
    \label{fig:core}
\end{figure}

To further investigate the inner regions' growth, we examine the relationship between core ($r<1$ kpc) stellar mass surface density (\(\Sigma_{*,1\rm{kpc}}\)) and total stellar mass (Figure \ref{fig:core}). The \(\Sigma_{*,1\mathrm{kpc}}\) is calculated by dividing the total stellar mass within a 1 kpc radius by the area enclosed within that radius. In Figure \ref{fig:core}, colored points represent the most massive galaxies described in Section \ref{sec:progen}, color-coded by redshift, while gray points show the full sample. Colored squares indicate the median values of \(\Sigma_{*,1\mathrm{kpc}}\) and total stellar mass at each redshift, with error bars showing the scatter in the data, estimated through bootstrapping. We summarize the median and error values in Table \ref{tab:comov}. The black, blue, and red lines indicate the fitted relations for the full sample, most massive galaxies sample (Section \ref{sec:progen}), and empirical relations from previous studies \citep[e.g.,][]{2015Sci...348..314T, 2013ApJ...776...63F}, respectively. 

Our sample (black and blue lines) follows a relatively shallow \(\Sigma_{*,1\rm{kpc}}\)–\(M_*\) relation compared to the local quiescent reference (red lines). At \(\log(M_*/M_\odot) \sim 11.5\), the core densities are already comparable to those of local quiescent galaxies \citep{2013ApJ...776...63F, 2015Sci...348..314T}, indicating that their inner structures closely resemble the local population \citep[e.g.,][]{2024ApJ...976...36Z, 2017ApJ...840...47B}. This suggests that the cores remain largely unchanged, with most structural evolution likely occurring in the outer regions.

Taken together, these findings suggest that these galaxies originated from compact, quenched structures that formed early in cosmic history \citep[e.g.,][]{2017ApJ...840...47B}. Over time, these galaxies have expanded in size as they accreted stellar mass in their outer regions through various mechanisms, most likely mergers \citep[e.g.,][]{Newman_2012, 2012ApJ...744...63O, 2016ApJ...828...27N}. These findings support a recent study by \citet{2024ApJ...971...99C} which argues that quenching at later cosmic times occurs through a more diverse set of pathways, while at higher redshifts, the quenching mechanisms more commonly produce compact remnants.

\subsection{Star Formation and Quenching Activities}

In Section \ref{sec:UVJ}, we show that the quenching histories of different regions within massive quiescent galaxies appear to be different. The \textit{UVJ} colors of these galaxies reveal that their inner regions tend to have quenched with shorter timescales compared to their outer regions. This difference in quenching timescales suggests that the dominant quenching mechanisms may vary across different parts of the galaxy.  

\subsubsection{Fast Quenching in the Inner Regions}

Our results indicate that the inner regions of massive galaxies experience a more rapid quenching process compared to their outer regions. Specifically, these inner regions appear to have been quenched at much earlier epochs, as they follow the rapid quenching track in the \textit{UVJ} diagram (Figure \ref{fig:UVJfitted}). This suggests that star formation in the cores was truncated quickly, likely due to powerful mechanisms acting at early times.

One plausible explanation for this early and efficient quenching is that the deep gravitational potential in the inner regions may accelerate the depletion of these gas reservoirs, effectively halting further star formation on short timescales \citep[e.g.,][]{2025MNRAS.536.2324L}. Additionally, another possible cause is the role of active galactic nucleus (AGN) feedback. The energy output from central black holes can drive both radiative and kinetic feedback, which heats or expels the cold gas needed for star formation \citep[e.g.,][]{2017FrASS...4...10C, 2022ApJ...941..205M}. 

To investigate this possibility, we cross-match our galaxy sample with several X-ray and quasar catalogs \citep[e.g.,][]{2015ApJS..220...10N, 2016ApJ...817...34M, 2018ApJS..236...48K, 2020ApJS..250....8L, 2023OJAp....6E..49F}.  {These catalogs primarily identify AGNs through X-ray emission or via spectroscopic signatures such as broad emission lines.} Based on these catalogs, we identify {{29} out of 256 galaxies ($\sim12\%$)} as quasars or X-ray sources. Most sources at $z > 2$ exhibit 2–10 keV X-ray luminosities exceeding $10^{43}$ erg/s (with only one reaching above $10^{44}$ erg/s), indicative of moderate to luminous AGN activity. {This early and rapid quenching phase, whether driven by AGN feedback or other processes, not only terminates star formation but also produces a compact and mature stellar population in the cores \citep[e.g., see][]{2024ApJ...971...99C}, setting the stage for the later, more gradual evolution observed in the outer regions}. 

{To quantitatively assess the possible impact of AGN emission on our measurements, we performed image decomposition of the X-ray confirmed sources using GALIGHT \citep{2020ApJ...888...37D}, modeling each galaxy with a Sérsic profile and a central point source. The possible point source contributes less than 30\% of the total flux in most cases, with only four sources exceeding this threshold. To test the significance of the point source, we compared GALIGHT fits with and without it. While the addition of a point source does not yield a significant improvement in chi-squared or residuals, the Sérsic + point-source models generally return lower chi-squared values, suggesting that a point source may be required but not strongly constrained. We then subtracted the fitted point source and re-performed the spatially resolved SED fitting with \pixedfit. Differences in the stellar mass and SFR radial profiles are generally small—central stellar masses decrease by about 0.1 dex and SFRs vary across radii—but the majority of these changes remain within the 1$\sigma$ uncertainties of the original results. We therefore conclude that while AGN activity may be present in a subset of our galaxies and could contribute to the early quenching of star formation in the cores, the AGN light itself does not significantly affect our photometry or bias the derived stellar masses and SFRs. Given this minimal impact and the uncertainty of the contribution from the possible AGN component, we do not apply any corrections for AGN contamination in our later discussions.}

\subsubsection{Delayed and Extended Quenching in the Outer Regions}

In contrast to the inner regions, the outer regions of these galaxies appear to have been quenched later and continue to grow, as indicated by their ongoing star formation activity observed at higher redshifts (Figures \ref{fig:profiles} and \ref{fig:UVJfitted}) and their gradual mass growth evolution (Figure \ref{fig:SFR_cum_mass}). At \( z \sim 3 \), these outskirts still exhibit signs of star formation, suggesting that the stellar mass in the outer regions was accumulated over an extended period rather than in a single rapid burst \citep[e.g.,][]{2017IAUS..321..327T}.  

A plausible explanation for this prolonged star formation is gas inflow triggered by mergers and interactions. As galaxies undergo minor or major mergers, cold gas can be funneled into their outskirts, sustaining star formation long after the central regions have quenched (e.g., \citealt{2015Sci...348..314T, 2007A&A...468...61D}). A recent simulation study by \cite{2025ApJ...982...30R} shows that some massive galaxies ($\log(M_*/M_\odot) > 5\times10^{10}$) that quenched by $z\sim3.4$ experience rejuvenated star formation by $z\sim2$ due to accreted gas, forming new stars between $r_{1/2}$ and $3r_{1/2}$, where $r_{1/2}$ is the half-mass radius. 

Accreted gas from mergers is expected to arrive from random orientations, rather than being organized into coherent, disk-like structures. To test this scenario, we analyze the images of our progenitor-descendant sample in the highest redshift bins. The radial profiles (Figure \ref{fig:profiles}) and rest-frame \textit{UVJ} colors (Figure \ref{fig:UVJfitted}) suggest elevated star formation activity at \( r \sim 5 \) kpc. {However, no disk-like structures are detected at these radii for most of our samples}. The absence of extended star-forming disks supports the hypothesis that star formation may have been triggered by gas accretion occurring from random orientations, likely driven by mergers or chaotic inflows \citep[e.g.,][]{2015MNRAS.450.2327Z, 2005A&A...437...69B}.  

Despite this sustained activity, the star formation activity of the outer region eventually decreases with decreasing redshift, with a longer quenching timescale than the inner regions. The gradual suppression of star formation may result from a combination of environmental processes, such as strangulation (halting fresh gas supply) and depletion of cold gas reservoirs \citep[e.g.,][]{2015Natur.521..192P}, alongside large-scale feedback mechanisms (e.g., AGN-driven winds) that indirectly limit gas availability \citep[e.g.,][]{2005Natur.433..604D}. 

\subsection{Merger Scenario} 

The clear disparity between the total mass growth rate and the observed SFR indicates that external processes, most likely mergers, rather than internal star formation, are responsible for most of the stellar mass accumulation. In this section, we explore the nature of these mergers, distinguishing between dry and wet mergers as well as minor and major merger events.

\subsubsection{Dry or Wet Mergers?}\label{sec:dry_wet}

Galaxy mergers are broadly classified by the gas content of the progenitor systems. In wet mergers, two gas-rich galaxies coalesce, and the abundant cold gas is compressed during the interaction, often triggering enhanced star formation (i.e., merger-induced starbursts) that can temporarily elevate the SFR of the remnant \citep{2008ApJ...681..232L}. In contrast, dry mergers occur between gas-poor systems—typically red, early-type galaxies—with little available cold gas. As a result, these mergers do not incite significant new star formation, and the stellar mass is built up primarily by the accretion of pre-formed stars \citep{2008ApJ...681..232L}.

In the sample from Section \ref{sec:progen}, our analysis reveals that in-situ star formation accounts for only about {8\% of the growth in the outer regions and 36\%} of the total stellar mass increase. {Moreover, the inferred total specific mass growth rate is up to 5 times higher than the estimated sSFR}. Such a substantial discrepancy implies that additional stellar mass must be deposited via external processes rather than being formed locally. Because wet mergers are typically accompanied by strong bursts of star formation \citep[e.g.,][]{1991ApJ...370L..65B, 1996ApJ...464..641M}, and our sample selection may miss such starburst galaxies, it suggests that the dominant mass assembly mechanism in these regions is through dry, gas-poor mergers.

Dry mergers, by their nature, involve the merging of systems that already have a high stellar-mass-to-gas mass ratio. As a result, these events contribute a substantial number of pre-formed stars to the outer envelopes of the remnant galaxy, without triggering the elevated star formation rates that would accompany gas-rich mergers. This scenario is consistent with previous studies showing that the late-time growth of massive galaxies is predominantly driven by mergers that do not trigger additional star formation \citep[e.g.,][]{2006ApJ...640..241B, 2010ApJ...709.1018V}. However, we note that the presence of prolonged star formation activities in the outer regions suggests that a small amount of accreted cool gas is still required.

\begin{figure}
    \centering
    \includegraphics[width=\linewidth]{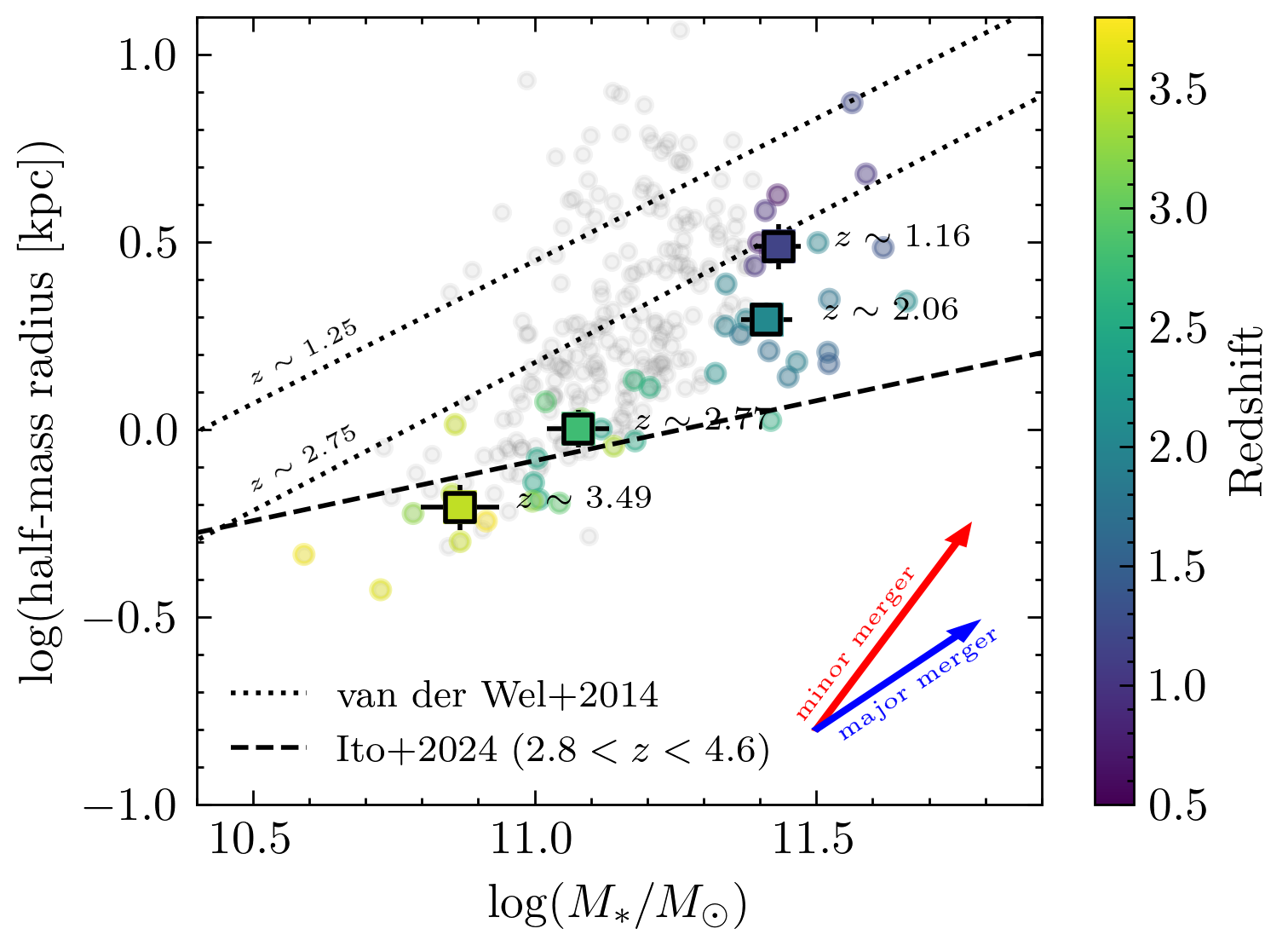}
    \caption{Grey points show the full sample, while colored points (coded by redshift) denote the selected subsample (Section \ref{sec:progen}). The red and blue arrows indicate the expected size–mass evolution for galaxies undergoing minor and major mergers, respectively. The dotted and dashed lines represent the results from previous studies by \citet{vanderWel_2014} and \citet{2024ApJ...964..192I}, respectively.}
    \label{fig:mass_rad}
\end{figure}

\subsubsection{Major vs. Minor Mergers}

{Galaxy mergers are commonly categorized as major or minor based on the mass ratio of the merging systems. In major mergers, galaxies of comparable mass merge, and the remnant structure reflects an average of the progenitors, leading to a size–mass scaling of $R_e \propto M_*^\alpha$ with $\alpha \lesssim 1$ due to roughly linear growth in both mass and size \citep[e.g.,][]{2006MNRAS.369.1081B, 2009ApJ...697.1290B}. In contrast, minor mergers involve a massive galaxy accreting a much smaller satellite, whose stars are deposited predominantly in the outer regions. This results in a stronger increase in effective radius relative to stellar mass growth, yielding $\alpha \gtrsim 2$. The steep size growth arises from tidal stripping and energy redistribution, which place the satellite’s stars at large radii, "puffing up" the host galaxy’s outskirts while minimally affecting the central regions \citep[e.g.,][]{2013MNRAS.429.2924H, 2012ApJ...744...63O}.
}
To test these scenarios, we show the half-mass radius versus stellar mass across different redshift intervals (Figure \ref{fig:mass_rad}). The half-mass radius is obtained from \texttt{statmorph} (see Section \ref{sec:results}). In Figure \ref{fig:mass_rad}, the colored points represent our selected sample from Section \ref{sec:progen}, color-coded by redshift, while the grey points indicate the full sample. Error bars indicate the scatter in the data, estimated through bootstrapping. The median and error values are summarized in Table \ref{tab:comov}. We also show the effective radius–stellar mass relations observed by \cite{vanderWel_2014} and \cite{2024ApJ...964..192I} as dotted and dashed lines, respectively. One of the most recent studies, \cite{2024ApJ...964L..10W}, reports that three of the most massive quiescent galaxies in their sample ($10.74 < \log(M_*/M_\odot) < 10.95$) at $3 < z < 5$ have rest-frame optical sizes of $\sim0.7$ kpc, comparable to our median half-mass radius of $\sim0.62\pm0.09$ kpc with $\log(M_*/M_\odot)\sim10.87\pm0.07$ at $z\sim3.5$.

{We fit the $R_e\propto M_*^\alpha$ relation to the medians of the colored points (shown as colored squares). For $z \lessapprox 2$, we find a size–stellar mass relation of $R_e \propto M_*^{2.67^{+1.14}_{-1.17}}$, which is broadly consistent with growth driven by minor mergers. This result aligns with many previous studies \citep[e.g.,]{2009ApJ...697.1290B, 2010ApJ...709.1018V, 2013MNRAS.429.2924H, 2023ApJ...956L..42S, 2024MNRAS.532.3604C}.}

{However, at $z \gtrapprox 2$, we find a flatter relation of $R_e \propto M_*^{0.91^{+0.20}_{-0.16}}$, which is more consistent with growth via major mergers. This apparent transition around $z \sim 2$ is also reflected in the stellar mass growth trend of our sample (Figure \ref{fig:mass_evol_progen}), where the total stellar mass shows little further increase at lower redshifts.}

{Given that major mergers can have relatively long timescales ($\sim$0.6 Gyr) \citep[see, e.g.,][]{2009MNRAS.399L..16C}, it is possible that these high-redshift galaxies experienced a major merger in the past, and their stellar mass distributions are still shaped by its aftermath. That said, we do not observe clear morphological signs of major mergers in the images of galaxies at $z > 2$. Additionally, we caution that this trend may be affected by our progenitor–descendant selection method. In particular, our lower-redshift sample may include newly quenched galaxies that are more compact than typical quiescent galaxies at the same redshift \citep[e.g.,][]{2017MNRAS.472.1401A}, rather than being the true descendants of the $z \sim 4$ quiescent population.}

\section{Conclusion} \label{sec:conclusion}

In this study, we conducted spatially resolved SED fitting with \pixedfit\ on a sample of rest-frame \textit{UVJ}-selected massive quiescent galaxies drawn from public JWST data in the CEERS and PRIMER surveys. Our main conclusions are as follows:
\begin{enumerate}
    \item The progenitors of massive quiescent galaxies in local universe began as compact, quenched systems. Subsequent mass accretion, primarily through mergers, has driven substantial size growth, especially in their outer regions.
\item \textit{UVJ} color analysis indicates that the inner regions quench rapidly, likely due to strong AGN feedback and the deep gravitational potentials of the cores. In contrast, the outer regions maintain star formation at higher redshifts and quench more gradually, possibly as a result of gas inflows from mergers.
\item The observed stellar mass growth in the outskirts, which exceeds the contribution from in situ star formation by up to 1 dex, strongly suggests that these galaxies have undergone mergers mostly with systems that have high stellar-to-gas mass ratios (i.e., dry mergers), although some cool gas are still required to explain the prolonged star formations.
\item {The central 1 kpc of most massive quiescent galaxies had already matured by $z\sim4$, showing little subsequent evolution. This supports a scenario in which bulge formation occurs early in their evolutionary history.}
\item {The most massive galaxies in our sample follow size-mass relation of \(R_e \propto M_*^{2.67^{+1.14}_{-1.17}}\) at $z\lessapprox2$, in agreement with the minor merger-driven evolution, and \(R_e \propto M_*^{0.91^{+0.20}_{-0.16}}\) at $z\gtrapprox2$, in agreement with the major merger-driven evolution.}
\end{enumerate}
Together, these findings support a two-phase evolutionary scenario in which massive quiescent galaxies form as compact systems at early times and subsequently expand via mergers.

\begin{acknowledgments}
{We thank the anonymous reviewer for their constructive feedback, which helped improve the quality of this work. This work is based on JWST observations obtained from the Mikulski Archive for Space Telescopes (MAST) at the Space Telescope Science Institute. The data used in this study are available at\dataset[doi:10.17909/f8zw-7h48]{https://doi.org/10.17909/f8zw-7h48}}. The data products presented herein were retrieved from the Dawn JWST Archive (DJA). DJA is an initiative of the Cosmic Dawn Center (DAWN), which is funded by the Danish National Research Foundation under grant DNRF140. The authors (Novan, Juan, and Ryo) gratefully acknowledge the support of the Japanese Government (Ministry of Education, Culture, Sports, Science and Technology or MEXT) scholarship for funding their studies. MA is supported by JSPS KAKENHI Grant-in-Aid for Scientific Research (B) Grant Number 24K00670. JPA gratefully acknowledges the support of the GPPU program at Tohoku University.
\end{acknowledgments}

%

\vspace{5mm}
\facilities{HST, JWST}


\software{\texttt{astropy} \citep{2013A&A...558A..33A,2018AJ....156..123A}, \pixedfit\ \citep{2021ApJS..254...15A}, Source Extractor \citep{1996A&AS..117..393B}, \texttt{statmorph} \citep{2019MNRAS.483.4140R}
          }




\bibliography{sample631}{}
\bibliographystyle{aasjournal}



\end{document}